\documentclass[12pt]{article} 
\usepackage{cite}
\usepackage{latexsym}
\usepackage{amsmath}
\usepackage{pstricks}
\newcommand{\ad}{^\dagger }
\newcommand{\becs}{\begin{cases}}
\newcommand{\bem}{\begin{matrix}}
\newcommand{\besp}{\begin{split}} 
\newcommand{\blp}{\bigl(} 
\newcommand{\brp}{\bigr)}
\newcommand{\ck}{\breve }
\newcommand{\coln}{\hbox{:}}

\newcommand{\encs}{\end{cases}}
\newcommand{\enm}{\end{matrix}}
\newcommand{\ensp}{\end{split}}

\newcommand{\lgl}{\langle }

\newcommand{\od}{\odot }
\newcommand{\ot}{\otimes }

\newcommand{\ra}{\rightarrow }
\newcommand{\Ra}{\Rightarrow }
\newcommand{\rgl}{\rangle }
\newcommand{\st}{\sqrt{2}}

\newcommand{\Tr}{{\rm Tr}}

\newcommand{\scriptl}[1] {{\cal #1}}

\newcommand{\FS}{\scriptl F }
\newcommand{\GS}{\scriptl G }
\newcommand{\HS}{\scriptl H }

\newcommand{\LS}{\scriptl L }

\newcommand{\al}{\alpha }

\newcommand{\lm}{\lambda }

\newcommand{\om}{\omega }

\newcommand{\amp}{&\hskip -.7em\relax }
\newcommand{\oda}{\amp\od\amp }
\newcommand{\vmt}{0.65 ex}

\begin{document}

\title{Consistent Resolution of Some Relativistic Quantum Paradoxes}

\author{Robert B. Griffiths
\thanks{Electronic mail: rgrif@cmu.edu}\\ 
Department of Physics,
Carnegie-Mellon University,\\
Pittsburgh, PA 15213, USA}


\date{Version of 2 July 2002}

\maketitle  

\begin{abstract}

	A relativistic version of the (consistent or decoherent) histories
approach to quantum theory is developed on the basis of earlier work by Hartle,
and used to discuss relativistic forms of the paradoxes of spherical wave
packet collapse, Bohm's formulation of Einstein-Podolsky-Rosen, and Hardy's
paradox.  It is argued that wave function collapse is not needed for
introducing probabilities into relativistic quantum mechanics, and in any case
should never be thought of as a physical process.  Alternative approaches to
stochastic time dependence can be used to construct a physical picture of the
measurement process that is less misleading than collapse models.  In
particular, one can employ a coarse-grained but fully quantum mechanical
description in which particles move along trajectories, with behavior under
Lorentz transformations the same as in classical relativistic physics, and
detectors are triggered by particles reaching them along such trajectories.
States entangled between spacelike separate regions are also legitimate quantum
descriptions, and can be consistently handled by the formalism presented here.
The paradoxes in question arise because of using modes of reasoning which,
while correct for classical physics, are inconsistent with the mathematical
structure of quantum theory, and are resolved (or tamed) by using a proper
quantum analysis.  In particular, there is no need to invoke, nor any evidence
for, mysterious long-range superluminal influences, and thus no
incompatibility, at least from this source, between relativity theory and
quantum mechanics.

\end{abstract}


	\section{Introduction}
\label{sct1}

	Three quarters of a century after the establishment of its basic
principles the physical interpretation of nonrelativistic quantum theory
remains a controversial subject.  The mathematical structure of the theory, a
suitable Hilbert space together with the unitary time evolution produced by
Schr\"odinger's equation, is universally accepted.  The controversy has to do
with the meaning to be assigned to a wave function, the role of measurements,
the significance of wave function collapse, the interpretation of macroscopic
quantum superpositions (Schr\"odinger's cat), a proper understanding of
entangled states --- such as in the famous Einstein-Podolsky-Rosen (EPR)
paradox --- and similar topics \cite{WhZr83}.  While a failure to understand
these matters has not prevented the application of quantum theory to an
enormous range of phenomena, it does make the subject confusing and difficult
for students, and for professional physicists who want to apply quantum
mechanics to a new domain, such as quantum information.  A good physical theory
requires both a sound mathematical framework and a consistent physical
interpretation,  and the latter is not entirely satisfactory in current quantum
mechanics textbooks.

	The situation does not improve upon going from nonrelativistic to
relativistic quantum mechanics and field theory.  The mathematics is more
elegant and harder to follow, but the same conceptual difficulties relating the
mathematics to physical reality remain; indeed, they are worse.  Wave function
collapse, which is something of an embarrassment for the nonrelativistic
theory, gives rise to serious conceptual problems in the relativistic case, and
there have been numerous discussions about this problem and how to deal with
it, among them
\cite{HlKr70,AhAl81,AhAl84,Flmn88,Vdmn93,Mdln94,ChHl95,Prs95c,Flmn96,%
Albr00,Ghrr00}.  The original EPR argument \cite{EnPR35} was formulated without
reference to relativity theory.  However, the fact that quantum theory predicts
violations of Bell's inequality \cite{Bll64,CHSH69}, together with the
experimental vindication of this prediction, \cite{AsDR82,KMWZ95}, is nowadays
often interpreted to mean that quantum mechanics is nonlocal in the sense that
certain causes can produce immediate effects a long (macroscopic) distance away
\cite{Mdln94,Stpp97}.  This, of course, calls into question a basic principle
of relativistic physics.  In addition there are other quantum paradoxes,
somewhat analogous to EPR, in which Lorentz invariance is an explicit part of
the construction \cite{Vdmn93,Ptws91,ClPP92,Hrdy92}, and their existence
suggests some conflict, or at least a certain tension, between quantum theory
and relativity.

	Most paradoxes of nonrelativistic quantum mechanics are closely linked
to a single fundamental difficulty which the founding fathers did not solve:
introducing probabilities into the theory in a fully consistent way.
Conventional textbook quantum theory, following the lead of von Neumann
\cite{vNmn32}, and London and Bauer\cite{LnBr39}, employs a deterministic
unitary time development based upon Schr\"odinger's equation, and then assumes
that a \emph{measurement} will, for some reason, have a random outcome whose
probability can be calculated, even though its existence cannot be justified,
using Schr\"odinger's wave function.  Assigning measurements this special role
in a fundamental theory seems rather odd, and generations of students have been
just as perplexed by it as were their teachers. To be sure, a bizarre idea that
helps organize our experience should not be rejected out of hand, and the
algorithm by which a wave function is used to calculate probabilities of
measurement outcomes has been extremely fruitful, with numerous results in very
good agreement with experiment. At the same time, the measurement approach has
given rise to an enormous set of conceptual headaches.  In the field of quantum
foundations these are referred to collectively as the \emph{measurement
problem}, and there has been very little progress in solving them
\cite{nnnmp}\vphantom{\cite{Wgnr63,Mttl98}}.  In short, while invoking
measurements makes it possible to calculate probabilities which agree with
experiment, in many other ways this approach to a fundamental understanding of
quantum theory causes more problems than it solves.

	In the last two decades methods based upon the idea of quantum
histories (consistent or decoherent histories) have been used to introduce
probabilities into quantum theory in a consistent way without making any
reference to measurement, by treating quantum dynamics as an inherently
stochastic process 
\cite{Grff84,GMHr90b,Omns92,GMHr93,Omns94,Grff96,Grff98,Omns99,Grff02}.  
This allows measurements to be thought of not as something
special, but as particular instances of quantum processes to which quantum
theory assigns probabilities using the same laws which apply to all other
processes. The probabilities of measurement outcomes obtained in this way are
identical with those computed using the older approach, and thus in complete
agreement with experiment.  But in the new approach measurements are no longer
necessary for interpreting quantum theory, and as a consequence the measurement
problem disappears.  This does not mean that quantum mechanics reduces to
classical physics. Instead, its seeming oddities, when properly understood, are
seen to be the consequences of a perfectly consistent mathematical and logical
structure, applicable to both microscopic and macroscopic systems, which
differs in crucial respects from that of classical physics.  In brief, quantum
reality is different from classical reality, just as relativistic reality
differs from (pre-relativistic) classical reality.

	Introducing probabilities in a consistent way without appealing to
measurements makes it possible to resolve or at least \emph{tame} the paradoxes
of nonrelativistic quantum theory, as shown in detail in Chs.~20 through 25 of
\cite{Grff02}.  The notion of taming a paradox can be illustrated by reference
to the well-known twin paradox of relativity theory.  Intuitively it seems
surprising that the astronaut who has been traveling for many years at high
speed returns to earth biologically much younger than his stay-at-home twin
brother.  But (special) relativity provides a consistent framework which allows
us to understand, in both mathematical and physical terms, why this can be so.
This explanation does not, and should not, remove our surprise when we first
encounter the difference between the relativistic idea of time and the notion
of absolute time that seems much closer to our everyday experience.  However,
once we understand relativistic principles the twin paradox is no longer a
conceptual headache, an unsolved mystery that calls into question our
understanding of physical reality.  Instead, it is a striking illustration of
how that reality differs from what we naively expected before studying it more
closely.

	The goal of the present paper is to apply the same approach,
probabilities not based on measurement, to relativistic versions of the
nonrelativistic paradoxes which have been successfully tamed by this method, in
particular, to relativistic versions of wave function collapse, EPR, and
Hardy's paradox.  Before presenting a brief outline of the rest of the paper,
it is worth noting that there are numerous conceptual difficulties and
paradoxes of relativistic quantum mechanics and quantum field theory which are
not addressed in the present paper. While there is no need to list all of them,
one in particular is worth mentioning: the problem of \emph{microlocality}, to
be distinguished (or so we believe) from that of the \emph{macrolocality}
needed for discussing the paradoxes just mentioned.  Microlocality is
associated, at least intuitively, with the idea that relativistic quantum
particles cannot be well localized in regions with linear dimensions which are
too small, nor precisely localized, in a sense which would please a
mathematician, in any finite region.  For example, to take the physicist's
point of view, it does not make sense to think of an electron localized in a
region smaller than its Compton wavelength.  Microlocality comes up in
Newton-Wigner states, and in Hegerfeldt's results on nonlocalization; see
\cite{Rdhd95,Flmn00,Hgrf98} for some representative literature.  The present
paper contains no attempt to resolve the mysteries of microlocalization;
instead the strategy, as in \cite{AhAl81,AhAl84}, is to avoid them, by setting
up relativistic quantum histories using a coarse-grained length scale:
distances which, though not necessarily macroscopic, are always significantly
larger than the relevant Compton or other length scale which might limit the
notion of locality employed in Sec.~\ref{sct3b}.  The resulting formulation can
at best be a good approximation, but we believe it is still sufficient for
taming those paradoxes with which we are concerned, for they involve quantum
correlations which can exist over length scales of centimeters or even, in the
case of light, meters or kilometers.  Thus we take the attitude that the
problems and paradoxes of macrolocality can be separated from issues of
relativistic microlocality. Should this be false it would, needless to say,
call into question the main results of this paper.  (Note that the histories
approach has been applied to some microlocal problems by Omn\`es
\cite{Omns97c}.)

	In order to make the present work self-contained, Sec.~\ref{sct2}
contains a summary of the essential ideas of the nonrelativistic histories
approach as formulated in \cite{Grff02}, and a specific example is considered
in Sec.~\ref{sct2d}, to make the presentation a bit less abstract.  (Here, and
later, we omit the arguments needed to show that various families of histories
are consistent or inconsistent, as they are not needed in order to follow the
presentation.  A detailed discussion of consistency conditions and methods for
checking them will be found in Chs.~10 and 11 of \cite{Grff02}; a more compact
presentation is in \cite{Grff96}.) The formulation of relativistic histories
presented in Sec.~\ref{sct3} follows in the footsteps of earlier work by Hartle
\cite{Hrtl91c}.  Most of the ideas are not new, but the way in which they are
presented owes something to developments in the nonrelativistic theory during
the last decade.  There is one important difference between our approach and
Hartle's. He employed regions with a finite extent in the time direction,
whereas we use spacelike hypersurfaces which at each point in space are
instantaneous in time.  Given that the present formulation is, as explained in
the previous paragraph, coarse grained in space, there is no reason not to
think of it as (in some sense) coarse grained in time, so the difference with
Hartle's formulation may not be all that significant.  There is other work
\cite{Omns97c,Blnc91} which has made use of relativistic histories and it is,
we believe, consistent with the present formulation in so far as they overlap.

	The discussion of relativistic paradoxes begins in Sec.~\ref{sct4} with
the collapse of the wave function of a single particle emitted in a spherical
wave as the result of some nuclear decay.  This, or rather a one-dimensional
analog which serves to illustrate the main points, is treated in some detail,
for in resolving (or taming) this paradox one employs most of the ideas needed
to handle relativistic versions of the Einstein-Podolsky-Rosen (EPR) paradox as
formulated by Bohm, taken up in Sec.~\ref{sct5}, and a paradox due to Hardy,
considered in Sec.~\ref{sct6}.  Studying these three paradoxes suffices, we
believe, to expose the basic principles needed to tame other paradoxes of the
same general sort, the kind which tempt one to think that the quantum world is
inhabited by mysterious influences which can propagate at superluminal speeds.
One of our main conclusions is that there are no such influences; belief in
them seems to have arisen through confusion over the proper rules for reasoning
about the physical properties of quantum systems, that is, logical difficulties
which are essentially the same in both the nonrelativistic and the relativistic
theory, although relativity adds a few interesting twists.  Counterfactual
forms of relativistic paradoxes are, strictly speaking, outside the scope of
the present paper, because analyzing them requires a relativistic
generalization of the formulation of counterfactual reasoning in \cite{Grff99}
and Ch.~19 of \cite{Grff02}, and this is not yet available.

	A concluding Sec.~\ref{sct7} provides a summary both of the principles
of relativistic histories in Sec.~\ref{sct3} and of the lessons learned through
exploring and taming the paradoxes in Secs.~\ref{sct4} to \ref{sct6}.

	\section{Nonrelativistic Quantum Histories} 
\label{sct2}

	\subsection{Kinematics}
\label{sct2a}

	There are by now a number of treatments of the basic principles of
nonrelativistic quantum theory from a histories perspective
\cite{Omns92,Omns94,Grff96,Grff98,Omns99,Grff02}.  While these differ in some
details, the basic strategy is the same; in what follows we use the notation in
\cite{Grff02}, where the reader will find a detailed discussion of various
points which, of necessity, are treated in a summary fashion in the present
discussion.

	The histories approach starts with the idea, which goes back to von
Neumann, Sec.~III.5 of \cite{vNmn32}, that any property of a quantum system at
a given instant of time corresponds to a \emph{subspace} of the quantum Hilbert
space, and the negation of this property to the orthogonal complement of this
subspace \cite{nnn0}.  Equivalently, a property is represented by a projector
$P$ (orthogonal projection operator) onto the subspace in question, and its
negation by the projector $I-P$, where $I$ is the identity operator.  Such
properties cannot, in general, be combined with one another in the manner which
is possible in classical physics.  For example, for a spin-half particle the
property that the $z$ component of angular momentum be positive, $S_z=+1/2$ in
units of $\hbar$, corresponds to a one-dimensional subspace in the Hilbert
space, as does its counterpart $S_x=+1/2$ for the $x$ component of angular
momentum.  In classical physics one would then be able to make sense of the
conjunction of these two properties: $S_z=+1/2$ AND $S_x=+1/2$.  But in quantum
theory this is not possible, at least without altering the rules of logic as
suggested by Birkhoff and von Neumann \cite{BrvN36}: $S_z=+1/2$ AND $S_x=+1/2$
is not a meaningful proposition, as it corresponds to no subspace in the
Hilbert space, nor is its negation $S_z=-1/2$ OR $S_x=-1/2$ a meaningful
proposition. (For more details, see \cite{Grff98,Grff02}.)  In the histories
approach two propositions which stand in such a relationship are called
\emph{incompatible}, and the basic strategy for avoiding the contradictions
associated with nonrelativistic quantum paradoxes is to insist that all valid
quantum descriptions consist of compatible entities: properties, histories,
etc.  In particular, properties corresponding to subspaces whose projectors do
not commute with each are always incompatible.
 
	A quantum \emph{history} is a sequence of quantum properties at a
succession of times, say
\begin{equation}
   t_0 < t_1 < t_2 < \cdots < t_f,
\label{eqn1}
\end{equation}
and has the form
\begin{equation}
  Y^\al = P^{\al_0}_0\od P^{\al_1}_1\od P^{\al_2}_2\od\cdots \od P^{\al_f}_f,
\label{eqn2}
\end{equation}
where $ P^{\al_j}_j$ is some projector representing a property of the system at
the time $t_j$.  The $\al_j$ is a label which differentiates this projector from other
projectors representing alternative properties which the system might possess
at this time.  The collection of such  projectors at time $t_j$ form a
\emph{decomposition  of the identity} $\{P^{\al_j}_j\}$, or
\begin{equation}
    I_j = \sum_{\al_j} P^{\al_j}_j. 
\label{eqn3}
\end{equation}
(The subscript on the identity operator $I$ can be ignored in the
nonrelativistic case, but is needed for the relativistic generalization.)  The
composite label $\al=(\al_0,\al_1,\al_2\ldots)$ on $Y$ in (\ref{eqn2})
identifies the history as a whole, and the collection of all histories of this
sort (for a fixed decomposition of the identity at each time) form a
\emph{sample space} of histories.  Note that the superscripts in (\ref{eqn2})
and (\ref{eqn3}) are labels, not powers. This usage need not cause any
confusion, since the square of a projector is the projector itself, and thus
there is never any need to raise it to some power.  One often considers
histories with a \emph{fixed initial state} of the form
\begin{equation}
  Y^\al =\Psi_0\od P^{\al_1}_1\od P^{\al_2}_2\od\cdots \od P^{\al_f}_f,
\label{eqn4}
\end{equation}
with $\Psi_0$ a single projector (possibly onto a pure state) independent of
$\al$.

	While the symbols $\od$ in (\ref{eqn2}) and (\ref{eqn4}) can be
regarded simply as spacers, equivalent to commas, it is actually convenient to
think of them as a variant of $\ot$, the operator for a
tensor product, so that $Y^\al$ is a projector on the Hilbert space
\begin{equation}
  \ck\HS = \HS_0\od\HS_1\od\HS_2 \cdots\od\HS_f
\label{eqn5}
\end{equation}
of histories, the tensor product of $f+1$ copies of the Hilbert space $\HS$ of
the system at a single time \cite{Ishm94}.  The product $Y^\al Y^{\bar\al}$ of
two projectors of the form (\ref{eqn2}) is zero if $\al\neq\bar\al$, that is,
if $\al_j$ is not equal to $\bar \al_j$ for some $j$.  Projectors on $\ck\HS$
of the form
\begin{equation}
  Y=\sum_\al \pi_\al Y^\al,
\label{eqn6}
\end{equation}
where each $\pi_\al$ is either 0 or 1, form a \emph{Boolean algebra} of history
projectors, all of which commute with one another.  This Boolean algebra, or
the sample space which generates it, is called a \emph{family} of histories,
and in the histories approach represents the \emph{event algebra} for a probability
theory.

	Whereas families of histories of the form (\ref{eqn2}) or (\ref{eqn4}),
with the projectors at any given time coming from a single decomposition
(\ref{eqn3}) of the identity, are the simplest kind to think about, the
histories formalism actually allows for much more general possibilities, see
Ch.~14 of \cite{Grff02}, which are sometimes useful. Since including these more
general families in a relativistic theory gives rise to no new problems or
issues, the exposition below and in Secs.~\ref{sct3} and \ref{sct4} is
restricted to the simpler type of family based on (\ref{eqn3}).  (A further
generalization allowed by Isham's formalism, projectors on the history space
(\ref{eqn5}) which cannot be written as a tensor product, as in (\ref{eqn3}),
or as a sum of projectors which are themselves tensor products, are excluded
from the present discussion, and from the relativistic generalization given
below.  Such histories have yet to be given any physical interpretation.)

	In ordinary probability theory one assumes that one and only one of the
mutually exclusive possibilities which make up a sample space (e.g., heads and
tails for a tossed coin) actually occurs.  Similarly, in the histories approach
to quantum theory, one
supposes that one and only one of the histories which make up the sample space
actually takes place in a given ``experimental run''.  In addition, if the
history $Y^\al$ for a given $\al$ is the one which actually occurs, the
successive projectors in (\ref{eqn2}) are thought of as representing actual
states of affairs at the times in question.  Thus the histories approach,
unlike textbook quantum theory, does not confine its physical interpretation to
measurements or the results of measurements.  Instead, measurements are
physical processes to be analyzed in the same way as all other physical
processes, by constructing appropriate histories of the total quantum system
including the measuring apparatus. This apparatus must be treated as a quantum
mechanical system, since the histories interpretation insists that everything
be discussed in quantum terms without introducing classical elements (except as
approximations to quantum theory).  In this way the histories approach
eliminates paradoxical elements of nonrelativistic quantum theory which arise
out of treating measurement as a fundamental concept.

	\subsection{Dynamics}
\label{sct2b}

	The time development of a quantum system, in the histories perspective,
is fundamentally a random or stochastic process, and the deterministic,
time-dependent Schr\"odinger equation is used as a tool to calculate the
probabilities of different histories.  (To be sure, the theory allows for
deterministic histories in which later events follow with probability one from
some initial condition.  But such ``unitary'' histories are exceptional cases;
most histories which are of interest in connection with actual laboratory
experiments are not of this form.)  As is well know, by integrating
Schr\"odinger's equation for a closed quantum system one can obtain a
collection of unitary \emph{time development operators}, denoted here by
$T(t',t)$, where the times $t'$ and $t$ serve as labels.  For a
time-independent Hamiltonian $H$ these operators can be written as
\begin{equation}
  T(t',t) = \exp[-i(t'-t) H/\hbar].
\label{eqn7}
\end{equation}
Whether or not $H$ depends on the time, as long as it is Hermitian the time
development operators satisfy the following conditions,

\begin{equation}
  T(t,t) =I,\quad T(t'',t') T(t',t) = T(t'',t),\quad 
  T(t',t) = T\ad(t,t') = T^{-1}(t,t').
\label{eqn8}
\end{equation}
for all $t$, $t'$, and $t''$.

	Given the time development operators, a \emph{chain} operator for the
history $Y^\al$ in (\ref{eqn2}) can be defined by writing its adjoint in the
form 
\begin{equation}
  K\ad(Y^\al) = P^{\al_0}_0 T(t_0,t_1)P^{\al_1}_1 T(t_1,t_2) 
P^{\al_2}_2 T(t_2,t_3)  \cdots T(t_{f-1},t_f) P^{\al_f}_f,
\label{eqn9}
\end{equation}
where the projectors appear in the same order as in \eqref{eqn9}; the operator
$K(Y^\al)$ is then a similar product with the operators in the reverse order,
and the arguments of each $T(t',t)$ interchanged. The \emph{weight} of
a history is given by
\begin{equation}
  W(Y^\al) = \lgl K(Y^\al),K(Y^\al)\rgl,
\label{eqn10}
\end{equation}
where the operator inner product $\lgl,\rgl$ is defined by
\begin{equation}
  \lgl A, B\rgl = \Tr(A\ad B),
\label{eqn11}
\end{equation}
assuming the trace exists.  In the case of a family of histories involving just
two times, $t_0$ and $t_1$, with an initial state $|\psi\rgl$ at $t_0$ and a
decomposition of the identity corresponding to an orthonormal basis
$|\phi^\al\rgl$ at $t_1$, the weights are given by
\begin{equation}
  W(\Psi_0\od P^\al_1) =| \lgl \phi^\al| T(t_1,t_0) |\psi\rgl |^2,
\label{eqn12}
\end{equation}
where $\Psi_0 = |\psi\rgl\lgl\psi|$ and $P^\al = |\phi^\al\rgl\lgl\phi^\al|$.
In this case the weights correspond to the usual Born transition probabilities,
and thus (\ref{eqn10}) can be thought of as a
generalization of the Born rule to the case of histories involving an arbitrary
number of times.

	The weights defined in \eqref{eqn10} can be combined with whatever
initial information one has about the quantum system in order to assign
probabilities to the various histories, in the same manner as for a classical
stochastic process; see Ch.~9 in \cite{Grff02}.  Thus, in particular, if the
system is known to have been in the initial state $|\psi\rgl$, the weight in
\eqref{eqn12} give the probabilities for the history $\Psi_0\od P^\al_1$ or,
equivalently, the probability that the quantum system will be in the state
$|\phi^\al\rgl$ at $t_1$, given that it was in the state $|\psi\rgl$ at $t_0$.

	When three or more times are involved, the histories
approach imposes additional conditions. In order that a family of
histories be acceptable as a possible stochastic description of a closed
quantum system, so that one can assign probabilities to the different
histories, the chain operators of the form
(\ref{eqn2}) must be \emph{mutually orthogonal},
\begin{equation}
  \lgl K(Y^\al), K(Y^{\bar\al})\rgl = 0 \text{ for } \al \neq \bar \al,
\label{eqn13}
\end{equation}
where $\lgl,\,\rgl$ is the operator inner product (\ref{eqn11}).  These are the
\emph{consistency conditions} or \emph{decoherence conditions}, and the left
side of \eqref{eqn13} is often referred to as a \emph{decoherence
functional}. (Various alternative consistency conditions have been proposed
from time to time; the one encountered most often is that in which the real
part of the operator inner product in \eqref{eqn17} rather than the product
itself is set equal to 0.)  A family of histories for which (\ref{eqn13}) is
satisfied is called a \emph{consistent family} or \emph{framework}.  A
meaningful description of a quantum system in physical terms is always based
upon some framework.

	The weights $W$ and the consistency conditions can also be expressed in
terms of Heisenberg projectors and chain operators, defined in the following
way.  Let $t_r$ be some \emph{reference time}; its value is unimportant as long
as it is held fixed.  Then for each projector entering a history of the form
(\ref{eqn4}), let the corresponding Heisenberg projector be defined by
\begin{equation}
  \hat P^{\al_j}_j = T(t_r,t_j) P^{\al_j}_j T(t_j,t_r).
\label{eqn14}
\end{equation}
The corresponding Heisenberg chain operator is
\begin{equation}
  \hat K\ad(Y^\al) = \hat P^{\al_0}_0 \hat P^{\al_1}_1\hat P^{\al_2}_2
  \cdots \hat P^{\al_f}_f,
\label{eqn15}
\end{equation}
which is (formally) simpler than (\ref{eqn9}) in that time development operators do
not appear on the right side.  It is then easy to check that (\ref{eqn10}) and
(\ref{eqn13}) are equivalent to 
\begin{equation}
  W(Y^\al) = \lgl \hat K(Y^\al),\hat K(Y^\al)\rgl,
\label{eqn16}
\end{equation}
\begin{equation}
  \lgl \hat K(Y^\al), \hat K(Y^{\bar\al})\rgl = 0 
   \text{ for } \al \neq \bar\al.
\label{eqn17}
\end{equation}

	Note that the operators on the right side on (\ref{eqn15}) do not (in
general) commute with each other, and hence the order is important.
Interchanging this order by using, for example, $\hat P^{\al_0}_0 \hat
P^{\al_2}_2\hat P^{\al_1}_1$ in place of $\hat P^{\al_0}_0 \hat P^{\al_1}_1\hat
P^{\al_2}_2$ for a history  based on the three times $t_0 < t_1 < t_2$
will (in general) change the value of the weight in (\ref{eqn16}).  Thus
keeping track of the temporal order of events is important if one wants to have
physically meaningful results. On the other hand, writing the projectors on the
right side of (\ref{eqn15}) in reverse order, with $\hat P^{\al_f}_f$ at the
left and $\hat P^{\al_0}_0$ at the right, merely replaces $\hat K\ad(Y^\al)$
with its adjoint $\hat K(Y^\al)$, and this does not alter $W(Y^\al)$ nor, if
the change is made for \emph{all} the histories in a family, does it alter the
consistency conditions (\ref{eqn17}).  Consequently, the histories
interpretation is invariant under a reversal of the direction of time. (Note
that this is quite a different issue from time-reversal invariance of the
Hamiltonian, which manifests itself in properties of the unitary operators
$T(t',t)$.)

	\subsection{Refinement and compatibility}
\label{sct2c}

	Let $\FS$ be a family of histories based upon decompositions of the
identity of the form (\ref{eqn3}) at a set of times given by (\ref{eqn1}).  We
shall say that a second family $\GS$ is a \emph{refinement} of $\FS$ (or $\FS$
is a \emph{coarsening} of $\GS$) provided two conditions are satisfied.

	1. The collection of times at which $\GS$ is defined includes all those
at which $\FS$ is defined, and perhaps some additional times.

	2. At each of the times for which $\FS$ is defined, $\GS$ is based upon
the same decomposition of the identity (\ref{eqn3}) as $\FS$, or else upon a
\emph{finer} decomposition of the identity, one in which at least one, and
possibly more, of the projectors in the original decomposition has been
replaced by two or more projectors which sum up to the projector which has been
replaced.

	These two conditions can be collapsed into a single condition if one
uses the following idea.  A history of the form (\ref{eqn2}) specifies certain
properties at the times given in (\ref{eqn1}), and says nothing about what is
happening at any other time.  Now one can ``extend'' the history (\ref{eqn2})
to additional times without changing its physical meaning if the identity $I$
is used as a projector for each added time, because $I$ represents the property
which is always true, and therefore its occurrence tells us nothing we did not
already know.  Given two families of histories which are not initially defined
at the same set of times, we can always extend the histories in the manner just
indicated so that we have equivalent families defined at a larger set of times,
which are now the same for both families.  If we allow for such an ``automatic
extension'', then $\GS$ is a refinement of $\FS$ if and only if at each time
where the histories in both families (of extended histories) are defined, the
decomposition of the identity for $\GS$ is the same or finer than that for
$\FS$.  Note that according to this definition, a family $\FS$ is always a
refinement of itself.  Also note that a refinement $\GS$ of a consistent family
$\FS$ may or may not be consistent. 

	Two frameworks $\FS$ and $\FS'$ are said to be \emph{compatible}
provided they possess a \emph{common refinement} which is
itself a consistent family or framework. That is, there must be some
family $\GS$ which is both a refinement of $\FS$ and a refinement of $\FS'$,
and which satisfies the consistency conditions.  Since according to the
definition given above, a family is always (formally) a refinement of itself,
$\GS$ could be $\FS$ or $\FS'$.  Indeed, if one framework is a refinement of
another, the two are compatible.  Frameworks which are not compatible are
called \emph{incompatible}.   
	There are two slightly different ways in which two frameworks can in
incompatible.  The first is that, as families, they have no common refinement:
this means that at at least one of the times of interest the two decompositions
of the identity contain projectors which do not commute with each other.  One
might call this ``kinematical incompatibility''.  But even if a common
refinement exists, it need not satisfy the consistency conditions,
leading to ``dynamical incompatibility.''

	A central principle of histories quantum theory is the \emph{single
framework rule} (or single family, or single set rule): a quantum description
must be constructed using a single consistent family, and results from two or
more incompatible frameworks cannot be combined.  This is an extension to
histories of the principle illustrated at the beginning of Sec.~\ref{sct2a}
using the $x$ and $z$ components of angular momentum of a spin-half particle,
and in the case of kinematic incompatibility can be justified on precisely the
same basis: the mathematics of the Hilbert space structure of quantum theory as
interpreted by von Neumann requires, if one takes it seriously as representing
physical reality, some changes in the way one thinks about that reality.
Incompatibility in this sense is a quantum concept that does not arise in
classical physics, and thus there is no good classical analogy for the single
framework rule.  Many paradoxes of nonrelativistic quantum theory involve some
violation of the single framework rule (see the discussion in Chs.~20 to 25 of
\cite{Grff02}), and the histories approach avoids these paradoxes by strictly
enforcing this rule, which plays an equally important role in relativistic
quantum theory.

	To complete this discussion, we note that when one uses the histories
approach, \emph{wave function collapse} is completely absent from the
fundamental principles of quantum theory.  If one treats quantum mechanics as a
stochastic theory, then various physical consequences can be worked out by
using the standard tools of probability theory, in particular, by computing
appropriate conditional probabilities.  For example, suppose that a measurement
has a probability 1/3 to turn out one way, apparatus pointer directed to the
left, and 2/3 to turn out a different way, pointer directed to the right.  If
the experiment is carried out and at the end the pointer points to the left,
then probability theory allows one to calculate various probabilities using
``pointer points to the left'' as a \emph{condition}.  Wave function collapse
as seen from a histories perspective provides a way (sometimes, but not always,
a useful way) to calculate certain conditional probabilities which can also be
computed by alternative methods.  In particular, wave function collapse is not
a mysterious physical phenomenon produced by an equally mysterious measurement
process.  One should think of it as something which occurs in the theoretical
physicist's notebook, not in the experimental physicist's laboratory! (In
addition, we shall sometimes use the term ``collapse'' in a metaphorical sense
to indicate the point at which a family of histories branches, as in
\eqref{eqn43}.)

	\subsection{Example using spin half}
\label{sct2d}

	It is helpful to see how the formalism described above applies to a
particular simple example, that of the spin degree of freedom of a spin-half
particle in zero magnetic field, so $T(t',t)=I$, the identity operator.
Suppose that the initial state at $t_0$ is $|z^+\rgl$ corresponding to
$S_z=+1/2$ in units of $\hbar$, and that at later times we use a decomposition
of the identity
\begin{equation}
  I = z^+ + z^-,
\label{eqn18}
\end{equation}
where $z^+$ is the projector $|z^+\rgl\lgl z^+|$, and $z^-$ the projector for
$S_z=-1/2$. Histories of the form \eqref{eqn4} based on the initial state 
$z^+$ then form a family
\begin{equation}
  \FS_0\coln\quad z^+\od \{z^+,z^-\} \od \{z^+,z^-\} \od \cdots,
\label{eqn19}
\end{equation}
in which each history begins with $z^+$, followed at later times by one of the
possibilities $z^+$ or $z^-$.  Because $T=I$, every history has zero weight or
zero probability, apart from $z^+\od z^+\od z^+\od\cdots$, which has
probability one.  In this example, and in many of those we will consider later,
most histories have zero weight and only a few occur with probability greater
than zero.  In such cases it is convenient to employ a shorthand in which
rather than listing all possible histories, as in \eqref{eqn19}, one shows only
those that have positive probability, the \emph{support} of the consistent
family.  In this shorthand \eqref{eqn19} is replaced with
\begin{equation}
  \FS_0\coln\quad z^+\od  z^+\od  z^+\od \cdots,
\label{eqn20}
\end{equation}
and there is no harm in referring to it as the ``framework $\FS_0$'' in place
of the more precise ``support of $\FS_0$''.  (While displaying the support is
usually adequate for indicating the family one has in mind, it does not
always determine unambiguously the decompositions of the identity, and
sometimes one has to be more specific about which histories of zero weight are
to included in the family.)  In the remainder of this paper we will use this
shorthand without further comment.

	The \emph{unitary} family \eqref{eqn20} in which each of the
projectors (in the support) is equal to its predecessor under the unitary map
produced by the time development operator is a rather special sort of quantum
description.  In practice one usually deals with \emph{stochastic} frameworks,
such as
\begin{equation}
  \FS_1\coln\quad z^+\od\left\{ \bem
 x^+\od x^+\od x^+\od \cdots,
\\[\vmt]
 x^-\od x^-\od x^-\od \cdots,
  \enm \right.
\label{eqn21}
\end{equation}
where $x^+$ and $x^-$ are projectors on the states $S_x=\pm 1/2$. The support
of this consistent family consists of two histories, both having the same
initial state, and each occurring with probability of 1/2.  In the first
history $S_z=+1/2$ at $t_0$ and $S_x=+1/2$ at $t_1$ and all later times.  It is
somewhat misleading to think of this history as one in which ``the spin is
pointing in the $z$ direction'' at $t_0$ and ``the spin is pointing in the $x$
direction'' at $t_1$ and later times, for this suggests that there is some
torque acting between $t_0$ and $t_1$ to make the spin precess, whereas we are
assuming there is no magnetic field present, and therefore no torque.  Instead,
the difference between $\FS_0$ and $\FS_1$ is that in the former one has chosen
to describe the $z$ component, and in the latter the $x$ component, of spin
angular momentum at times later than $t_0$.  A description of a classical
spinning object which specifies one component, say $L_z$ of its angular
momentum at an earlier time and a different component, say $L_x$, at a later
time tells one nothing about the direction of the total angular momentum at
either time, and this is a helpful analogy in thinking about the quantum case,
where $S_z=+1/2$ does \emph{not} imply that $S_x$ or $S_y$ is zero.

	In $\FS_1$ the two histories ``split'', or diverge from each other,
at $t_1$, but there are other frameworks in which this split occurs later, such
as 
\begin{equation}
  \FS_2\coln\quad z^+\od z^+\od \left\{ \bem
 x^+\od x^+\od \cdots,
\\[\vmt]
 x^-\od x^-\od \cdots,
  \enm \right.
\label{eqn22}
\end{equation}
where it occurs at $t_2$.  In view of the remarks in the previous paragraph, it
is evident that the presence as well as the timing of such a split --- one
could also call it a ``collapse'' --- is not some sort of physical effect.
Instead, it arises from the possibility of constructing various different,
incompatible (in the quantum sense) stochastic descriptions of the same quantum
system starting in the same initial state. This does \emph{not} mean that one
of these descriptions is correct and the others false, but rather that there is
no way of combining them into a single description.  This is obvious in the
case of $\FS_1$ and $\FS_2$ because at $t_1$ the former assigns a value to
$S_x$ and the latter a value to $S_z$, whereas the Hilbert space does not allow
simultaneous values of two different components of spin angular momentum.  In
the same way, both $\FS_1$ and $\FS_2$ are incompatible with $\FS_0$.  The
choice of which family to use in a particular circumstance is made by the
physicist on the basis of what aspects of the time development he wants to
discuss.  If it is $S_x$ at $t_2$, then either $\FS_2$ or $\FS_1$ can be used,
but not $\FS_0$, whereas neither $\FS_2$ nor $\FS_1$ can be used to describe
$S_z$ at $t_2$.  Also note that once a split or collapse of the kind one finds
in $\FS_1$ or $\FS_2$ has occurred, it cannot be undone by, for example,
replacing $x^+$ with $z^+$ at $t_3$ in both histories in \eqref{eqn22} (or in
\eqref{eqn21}). Such a family would violate the consistency conditions, and
hence not be a meaningful stochastic description of the time development of
this quantum system.

	One can extend this example to include measurements.  Let $|X\rgl$
represent the initial state of an apparatus designed to measure $S_x$, and
suppose that during the time interval the total system of particle plus
apparatus undergoes a unitary time evolution given by:
\begin{equation}
  |x^+\rgl\ot |X\rgl \ra |x^+\rgl\ot |X^+\rgl,\quad
 |x^-\rgl\ot |X\rgl \ra |x^-\rgl\ot |X^-\rgl.
\label{eqn23}
\end{equation}
Before and after this time both particle and apparatus remain unchanged. (One
can imagine that the particle passes through the apparatus between $t_1$ and
$t_2$, but that for simplicity we have omitted the center of mass motion of the
particle from our description.)  It is helpful to think of $|X^+\rgl$ and
$|X^-\rgl$ as macroscopically distinct apparatus states, e.g., corresponding to
two positions of a visible pointer.  This is an oversimplified but not
misleading description of a quantum measurement; see Sec.~17.4 of \cite{Grff02}
for a more realistic approach.
	(Typical laboratory measurements or quantum systems are destructive in
the sense that the measured property is significantly altered in the
measurement process.  The histories approach handles these without difficulty,
see Ch.~17 of \cite{Grff02}, but \eqref{eqn23} is a nondestructive model of
measurement, which makes it easier to compare with usual textbook approach.)

	Let us suppose that at $t_0$ the combined system is in a state 
$z^+\ot X$, i.e., $S_z=+1/2$ for the particle, and the apparatus in its
``ready'' state.  One possible framework is that of unitary time evolution of
the total system:
\begin{equation}
  \GS_0\coln\quad \Psi_0\od z^+ X\od S\od S\od\cdots,
\label{eqn24}
\end{equation}
where the initial state is
\begin{equation}
  \Psi_0 = z^+ X;
\label{eqn25}
\end{equation}
omitting the $\ot$ between the projectors $z^+$ and $X$ onto the states
$|z^+\rgl$ and $|X\rgl$ does not lead to any ambiguity.  The projector $S$
projects onto the state
\begin{equation}
  |S\rgl = \blp |x^+\rgl |X^+\rgl + |x^-\rgl |X^-\rgl\brp/\st,
\label{eqn26}
\end{equation}
which, since the apparatus is of macroscopic size, is a \emph{macroscopic
quantum superposition} (MQS) or Schr\"odinger cat state.  As a consequence,
$\GS_0$, even though a perfectly correct quantum description of the time
development, is not of much use for discussing the measurement process in
physical terms.  The reason is that $S$ does not commute with either of the
projectors $X^+$ or $X^-$ describing the possible measurement outcomes, so if
one uses the description provided by $\GS_0$ it is meaningless to ascribe a
position to the pointer after the measurement has taken place.

	Of greater utility is the framework
\begin{equation}
  \GS_1\coln\quad \Psi_0\od\left\{ \bem
 x^+X\od x^+X^+\od x^+X^+\od \cdots,
\\[\vmt]
 x^-X\od x^-X^-\od x^-X^-\od \cdots,
  \enm \right.
\label{eqn27}
\end{equation}
which is the measurement counterpart of $\FS_1$ in \eqref{eqn21}. In this
family the apparatus is in its ready state $X$ and the particle is in one of
the two states $S_x=\pm 1/2$ at $t_1$.  At $t_2$ and later times the state
$X^\pm$ of
the apparatus reflects the earlier state of the particle, as one would expect
given \eqref{eqn23}. From the measurement outcome $X^+$ at any time after
$t_2$, one can infer (conditional probability equal to 1) that $S_x=+1/2$ both
before and after the measurement; similarly $X^-$ implies $S_x=-1/2$ at earlier
as well as later times.

	The measurement counterpart of $\FS_2$ is the framework
\begin{equation}
  \GS_2\coln\quad \Psi_0\od z^+X\od\left\{ \bem
 x^+X\od x^+X^+\od \cdots,
\\[\vmt]
 x^-X\od x^-X^-\od \cdots.
  \enm \right.
\label{eqn28}
\end{equation}
It corresponds fairly closely to the traditional ``collapse'' picture of the
measurement process found in textbooks, since one has unitary time development
until the particle interacts with the apparatus, after which the particle
state, $x^+$ or $x^-$, is correlated with the measurement outcome state $X^+$
or $X^-$.  However, $\GS_2$ is only one of a collection of equally valid but
mutually incompatible ways of using quantum mechanics to describe the measuring
process.  From the point of view of fundamental quantum theory there is no
reason to prefer $\GS_2$ to the unitary family $\GS_0$.  To be sure, the latter
cannot be used to describe the measurement outcome, for, as pointed out
earlier, $S$ does not commute with $X^+$ or $X^-$.  Thus from a practical point
of view $\GS_2$ is more useful than $\GS_0$.  But there is no reason to prefer
$\GS_2$ to $\GS_1$, and $\GS_1$ has the advantage that it allows one to think
of the measurement process as a \emph{measurement} in the usual sense of that
term: a procedure by which the macroscopic outcome reflects a property the
measured system had \emph{before} the measurement takes place.  In practice,
most measurements on microscopic quantum systems carried out in the laboratory
can best be thought of using a viewpoint akin to that of $\GS_1$: a gamma ray
is detected by destroying it, the momentum of a charged particles emerging from
a collision vertex is measured by changing it in a magnetic field, etc.  (In
these cases \eqref{eqn23} is not an appropriate model, because the measurements
are destructive, but the histories approach handles these equally well, Ch.~17
of \cite{Grff02}, and shows that the measurement outcomes are correlated with
quantum states which existed before the measurement interaction.)  Descriptions
analogous to $\GS_2$ play very little role in physics apart from their
appearance in textbook lists of quantum axioms where they have confused
generations of students, not because they are wrong, but because the
corresponding ``wave function collapse'' has been misinterpreted as a physical
phenomenon, rather than just one of many ways of describing quantum time
development.  Rectifying that misinterpretation is, as we shall see, the key to
untangling several relativistic quantum paradoxes.

	\section{Relativistic Quantum Histories} 
\label{sct3}

	\subsection{Kinematics and dynamics}
\label{sct3a}

\begin{figure}[h]
$$
\begin{pspicture}(-1.0,-0.5)(5,5.0) 
\def\lwn{0.025} 
\def\lwd{0.05}  
\def\lwdsh{0.020} 
\def\txl{-0.2} 
\psset{
labelsep=2.0,
arrowsize=0.150 1,linewidth=\lwd}
\psline[linewidth=\lwn]{->}(-1.0,0)(5.0,0)
\psline[linewidth=\lwn]{->}(0,-0.5)(0,4.5)
\rput[l](0,4.7){$t$}
\rput[b](5.2,0){$x$}
\psline(-1.0,4.0)(5.0,3.0)
\psbezier(-1.0,3.0)(-0.5,3.2)(-0.2,3.1)(0.0,3.0)
\psline(0.0,3.0)(1.0,2.5)
\psbezier(1.0,2.5)(1.4,2.3)(2.5,2.0)(3.0,2.0)
\psline(3.0,2.0)(4.0,2.0)
\psbezier(4.0,2.0)(4.3,2.0)(4.7,1.8)(5.0,1.5)
\psline(-1.0,1.7)(5.0,0.7)
\psbezier(-1,-0.2)(-.7,-0.15)(-0.3,0.10)(0,0.25)
\psline(0,0.25)(1,0.75)
\psbezier(1,0.75)(2,1.25)(3,0.7)(5,-0.5)
\rput[r](\txl,0.4){$S_0$}
\rput[r](\txl,1.3){$S_1$}
\rput[r](\txl,2.8){$S_2$}
\rput[r](\txl,4.2){$S_3$}
\end{pspicture}
$$

\caption{%
A possible collection of time-ordered spacelike hypersurfaces.}
\label{fgr1}
\end{figure}
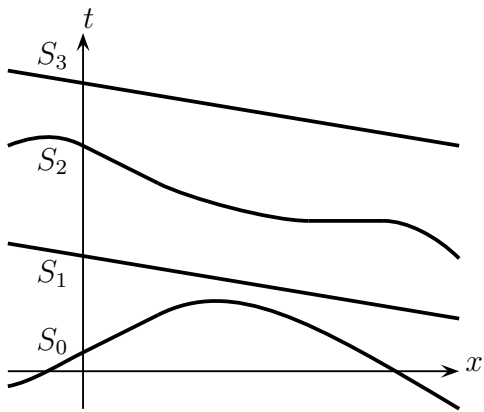

	A plausible generalization of the histories approach described in
Sec.~\ref{sct2} can be carried out in the following way.  Introduce a
collection $\{S_j\}$, $j=0,1,2,\ldots$, of smooth, infinite, nonintersecting
three-dimensional spacelike hypersurfaces, as suggested by the diagram in
Fig.~\ref{fgr1}.  They do not have to be ``flat'' hyperplanes, but the
requirement that no two surfaces intersect means that if two or more
hyperplanes belong to the collection, they must be parallel. As they do not
intersect,  the hypersurfaces can be ordered in time, and we assume that
$S_j$ is earlier than $S_{j+1}$, with $S_0$ the earliest hypersurface.
	(These spacelike surfaces have no thickness in the time direction,
unlike the open regions in space-time employed by Hartle \cite{Hrtl91c},
Blencowe \cite{Blnc91}, and in algebraic quantum field theory \cite{Hg92}.
This is consistent with our decision, Sec.~\ref{sct1}, to ignore problems of
microlocality.  Should it be necessary for technical mathematical reasons to
introduce a small but finite thickness or duration in the time direction, that
should not alter our conclusions.)

	Next assume that for each hypersurface $S_j$ there is a Hilbert space
$\HS_j$ with identity operator $I_j$.  Given a decomposition (\ref{eqn3}) of
$I_j$ in projectors, one can define histories of the form (\ref{eqn2}) on the
history Hilbert space (\ref{eqn5}).  The dynamical laws can be expressed using
a collection of time development operators $\{T_{jk}\}$, where $T_{jk}$ is a
unitary map (bijective isometry) from $\HS_k$ onto $\HS_j$, the
analog of the nonrelativistic $T(t_j,t_k)$.  The conditions analogous to
(\ref{eqn8}) are, obviously,
\begin{equation}
  T_{jj}=I_j, \quad T_{ij} T_{jk}=T_{ik}, \quad T\ad_{jk} = T_{kj}.
\label{eqn29}
\end{equation}

	At this point one could introduce chain operators of the form
\eqref{eqn9}, but for our purposes it is more convenient to introduce
Heisenberg projectors on the Hilbert space $\HS_r$ of a special \emph{reference
hypersurface} (or hyperplane) $S_r$.  As we shall ascribe no \emph{physical}
significance to the Heisenberg projectors --- they are only introduced as a
convenience for mathematical calculations --- the relationship between $S_r$
and the collection $\{S_j\}$ is arbitrary; in particular $S_r$ may intersect
the other hypersurfaces, or it could be identical to one of them.  By means of
the unitary time development operators $T_{jr}$ mapping $\HS_r$ to the other
Hilbert spaces we define the Heisenberg operator
\begin{equation}
  \hat P^{\al_j}_j = T_{rj} P^{\al_j}_j T_{jr}.
\label{eqn30}
\end{equation}
corresponding to the projector $P^{\al_j}_j$. Heisenberg chain operators
mapping $\HS_r$ to itself are then defined as a product of Heisenberg
projectors \eqref{eqn15}, and the weights and consistency conditions are
expressed in terms of these chain operators using (\ref{eqn16}) and
(\ref{eqn17}), with an appropriate definition of the operator inner product
$\lgl,\rgl$.

	Defining $\lgl,\rgl$ in terms of the trace, as in \eqref{eqn11}, is
only satisfactory if the trace exists, which need not be the case, since
$\HS_r$ is infinite.  There is no problem if all the histories we are
interested in are of the form \eqref{eqn4} with $\Psi_0$ a pure initial state
or a projector onto a finite subspace of $\HS_0$.  Alternatively, one can
introduce a density operator $\rho_0$ (with unit trace) on $\HS_0$, define its
Heisenberg counterpart $\hat \rho = T_{r0}\rho_0 T_{0r}$ as in \eqref{eqn30},
and replace the operator inner product $\lgl,\,\rgl$ in \eqref{eqn16} and
\eqref{eqn17} with
\begin{equation}
  \lgl A, B\rgl_\rho = \Tr(\hat \rho A\ad B).
\label{eqn31}
\end{equation}  
(In this case it is best to regard $\hat \rho$ as a pre-probability; see
Sec.~15.2 of \cite{Grff02}.)  Of course, any other $\HS_j$ could be used in
place of $\HS_0$, but physicists typically tend to employ an initial condition
(we live in a thermodynamically irreversible world).
	Given a family of histories satisfying the consistency conditions
\eqref{eqn17}, its physical interpretation is precisely the same as in the
nonrelativistic case: one and only one history belonging to the family actually
occurs in any given situation.  The probabilities of histories are determined
by the weights and whatever constitutes one's information about the initial
state or experimental setup; see, e.g., Sec.~9.1 of \cite{Grff02}. 

	Since histories with events on a finite set of spacelike hypersurfaces
may seem odd to a reader accustomed to the continuous time trajectories
familiar in classical physics, the following comments may be helpful.  Just as
in the nonrelativistic case --- see the discussion of refining a family in
Sec.~\ref{sct2c} --- it is always possible to introduce additional spacelike
hypersurfaces between (or before or after) those in the collection $\{S_j\}$,
and extend histories of the form (\ref{eqn2}), without changing their physical
meaning, by introducing the trivial event $I$ on these additional
hypersurfaces.  This shows that defining histories on a finite collection of
hypersurfaces does not imply that the world ceases to exist at intermediate
space-time points, it simply means that these histories contain no information
about what is happening elsewhere than on these hypersurfaces.  Think of being
outside on a dark night during a thunder storm, when flashes of lightening
illuminate the landscape at certain times, but nothing can be seen in the
intervening intervals.
	In the nonrelativistic case one can, to be sure, produce histories
which are described by non-trivial (not equal to $I$) projectors at all times,
thus ``filling in the gaps'' in (\ref{eqn2}); one method of doing this is
discussed in Sec.~11.7 of \cite{Grff02}. However, different ways of filling the
gaps lead to incompatible families, and since there is no limit to the
number of times which enter a discrete history of the form (\ref{eqn2}),
there is really no need to fill the gaps from the point of view of
providing an adequate physical description.  The same comment applies in the
relativistic case, at least for the purposes of the present paper. 

	Just as in Sec.~\ref{sct2}, a family of histories satisfying the
consistency conditions will be called a \emph{consistent family} or
\emph{framework}.  A refinement $\FS'$ of a framework $\FS$ must include among
its hypersurfaces $\{S'_k\}$ all the hypersurfaces associated with $\FS$, and
on the latter the decomposition of the identity used in $\FS'$ must be a
refinement of the one used in $\FS$.  In order for it to be a framework, $\FS'$
must satisfy the consistency conditions.
	Two frameworks $\FS$ and $\GS$ will be said to be \emph{compatible}
provided they possess a common refinement which is itself a framework;
otherwise they are \emph{incompatible}.  This is the same definition employed
in the nonrelativistic case. The \emph{single framework rule} is also the same
as for nonrelativistic quantum theory: quantum descriptions must always be
constructed using a single framework.  If two frameworks are not identical but
are compatible, a common description can be constructed using their common
refinement.  However, descriptions corresponding to incompatible frameworks
cannot be combined.

	\subsection{Local regions and properties}
\label{sct3b}

\begin{figure}[h]
$$
\begin{pspicture}(-4.5,-0.8)(8.5,4.8) 
\def\lwn{0.025} 
\def\lwd{0.04}  
\def\lwdsh{0.05} 
\def\txl{-0.2} 
\def\txlp{0.5} 
\def\txr{3.5} 
\psset{
labelsep=2.0,
arrowsize=0.150 1,linewidth=\lwd}
		\def\figa{
\psline[linewidth=\lwn]{->}(-1.0,0)(5.0,0)
\psline[linewidth=\lwn]{->}(0,-0.5)(0,4.5)
\rput[l](0,4.7){$t$}
\rput[b](5.2,0){$x$}
\psline[linestyle=dotted,linewidth=\lwdsh](-1,4)(0,3.833)
\psline(0,3.833)(1,3.667)
\psline[linestyle=dotted,linewidth=\lwdsh](1,3.667)(3,3.333)
\psline(3,3.333)(4,3.167)
\psline[linestyle=dotted,linewidth=\lwdsh](4,3.167)(5,3.0)
\psbezier[linestyle=dotted,linewidth=\lwdsh](-1.0,3.0)(-0.5,3.2)%
(-0.2,3.1)(0.0,3.0)
\psline(0.0,3.0)(1.0,2.5)
\psbezier[linestyle=dotted,linewidth=\lwdsh](1.0,2.5)(1.4,2.3)%
(2.5,2.0)(3.0,2.0)
\psline(3.0,2.0)(4.0,2.0)
\psbezier[linestyle=dotted,linewidth=\lwdsh](4.0,2.0)(4.3,2.0)%
(4.7,1.8)(5.0,1.5)
\psline[linestyle=dotted,linewidth=\lwdsh](-1,1.7)(3,1.033)
\psline(3,1.033)(4,0.867)
\psline[linestyle=dotted,linewidth=\lwdsh](4,0.867)(5,0.7)
\psbezier[linestyle=dotted,linewidth=\lwdsh](-1,-0.2)(-.7,-0.15)%
(-0.3,0.10)(0,0.25)
\psline(0,0.25)(1,0.75)
\psbezier[linestyle=dotted,linewidth=\lwdsh](1,0.75)(2,1.25)%
(3,0.7)(5,-0.5)
\rput[r](\txl,0.4){$S_0$}
\rput[r](\txl,1.3){$S_1$}
\rput[r](\txl,2.8){$S_2$}
\rput[r](\txl,4.2){$S_3$}
\rput[b](\txlp,0.7){$R_0$}
\rput[b](\txlp,2.9){$R'_2$}
\rput[b](\txlp,3.9){$R'_3$}
\rput[b](\txr,1.1){$R_1$}
\rput[b](\txr,2.1){$R''_2$}
\rput[b](\txr,3.4){$R''_3$}
\rput(2.0,-0.5){\textbf{(a)}}
} 
		\def\figb{
\psline[linewidth=\lwn]{->}(-1.0,0)(5.0,0)
\psline[linewidth=\lwn]{->}(0,-0.5)(0,4.5)
\rput[l](0,4.7){$t$}
\rput[b](5.2,0){$x$}
\psline[linestyle=dotted,linewidth=\lwdsh](-1,4)(0,3.833)
\psline(0,3.833)(1,3.667)
\psbezier[linestyle=dotted,linewidth=\lwdsh](1,3.667)(2,3.5)
(3,3.833)(5,3.5)
\psbezier[linestyle=dotted,linewidth=\lwdsh](-1,3.5)(1,3.167)%
(2,3.5)(3,3.333)
\psline(3,3.333)(4,3.167)
\psline[linestyle=dotted,linewidth=\lwdsh](4,3.167)(5,3.0)
\psbezier[linestyle=dotted,linewidth=\lwdsh](-1.0,3.0)(-0.5,3.2)%
(-0.2,3.1)(0.0,3.0)
\psline(0.0,3.0)(1.0,2.5)
\psbezier[linestyle=dotted,linewidth=\lwdsh](1.0,2.5)(2,2.0)%
(4,2.5)(5,2.0)
\psbezier[linestyle=dotted,linewidth=\lwdsh](-1.0,2.5)(0,2.0)%
(2,2.0)(3.0,2.0)
\psline(3.0,2.0)(4.0,2.0)
\psbezier[linestyle=dotted,linewidth=\lwdsh](4.0,2.0)(4.3,2.0)%
(4.7,1.8)(5.0,1.5)
\psbezier[linestyle=dotted,linewidth=\lwdsh](-1,-0.2)(-.7,-0.15)%
(-0.3,0.10)(0,0.25)
\psline(0,0.25)(1,0.75)
\psbezier[linestyle=dotted,linewidth=\lwdsh](1.0,0.75)(2.5,1.5)%
(4,1.4)(5,1.0)
\psbezier[linestyle=dotted,linewidth=\lwdsh](-1.0,-0.5)(2,1.0)%
(2,1,2)(3,1.033)
\psline(3,1.033)(4,0.867)
\psline[linestyle=dotted,linewidth=\lwdsh](4,0.867)(5,0.7)
\rput[b](\txlp,0.7){$R_0$}
\rput[b](\txlp,2.9){$R'_2$}
\rput[b](\txlp,3.9){$R'_3$}
\rput[t](\txr,0.9){$R_1$}
\rput[t](\txr,1.9){$R''_2$}
\rput[t](\txr,3.1){$R''_3$}
\rput(2.0,-0.5){\textbf{(b)}}
} 
\rput(-3.5,0){\figa}
\rput(3.5,0){\figb}
\end{pspicture}
$$
\caption{%
(a) Finite regions $R_j$, possibly consisting of more than one connected piece,
belonging to infinite spacelike hypersurfaces. (b) An alternative way of
embedding the same finite regions in spacelike hypersurfaces.}
\label{fgr2}
\end{figure}
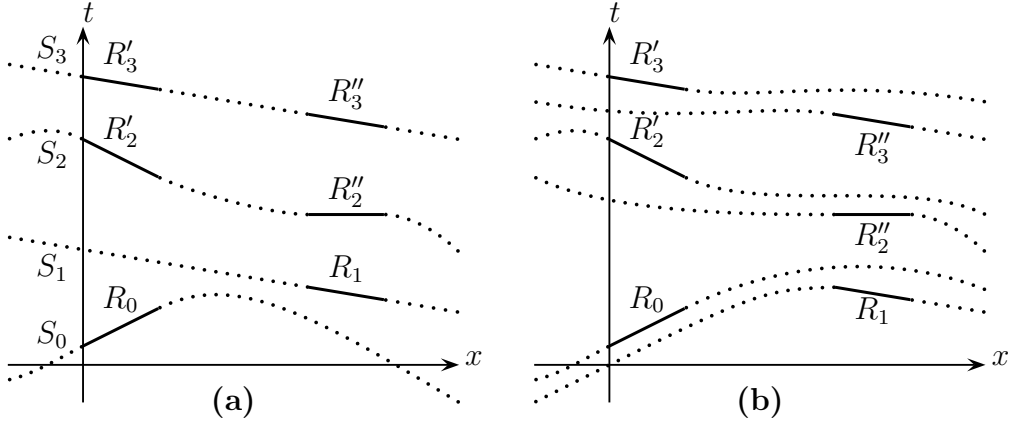

	For discussing macrolocality and quantum paradoxes we shall want to
consider spacelike regions $R_j$ of finite extent, see Fig.~\ref{fgr2}(a), each
consisting of one or else a small number of connected pieces belonging to a
spacelike hypersurface $S_j$, with each piece of ``macroscopic'' size, much
larger than a Compton wavelength, with ``reasonable'' (e.g., piecewise smooth)
boundaries --- imagine a sphere or a cube.  (Much of the following discussion
is valid if $R_j$ is an infinite piece of $S_j$, but we shall be interested in
cases in which it is finite.)  By making each $R_j$ part of some $S_j$, with
the collection $\{S_j\}$ satisfying the conditions given in Sec.~\ref{sct3a},
we ensure that it is possible to impose a well-defined \emph{time ordering} on
these finite regions.  One could also do this by working out a set of
conditions applicable directly to the collection $\{R_j\}$, but requiring that
they be embeddable in a time ordered collection of infinite hypersurfaces is a
fairly simply and intuitive way to proceed.

	The lowbrow way to think of a property $P_j$ as local to or localized
in $R_j$ is to imagine that the Hilbert space $\HS_j$ associated with $S_j$ is
a tensor product $\HS^r_j\ot\HS^s_j$, with $\HS^r_j$ associated with $R_j$ and
$\HS^s_j$ associated with the complement of $R_j$ in $S_j$.  Then suppose that
on this tensor product $P_j$ is of the form $P^r_j\ot I^s_j$, with $I^s_j$ the
identity on $\HS^s_j$.  Thus $P^r_j$, an operator on $\HS^r_j$, tells one
something about the state of affairs inside $R_j$, while $I^s_j$ is totally
uninformative about what is going on elsewhere.  Note that the usual
physicists' convention allows the same symbol $P_j$ to represent $P^r_j$ or
$P^r_j\ot I^s_j$, without (much) risk of confusion, and we shall make use of
this liberty.  The highbrow way of thinking about a localized property requires
dealing seriously with the microlocality problem, see Sec.~\ref{sct1}, and is
outside the scope of the present paper.
	If $R_j$ can be embedded in two different spacelike hypersurfaces $S_j$
and $S'_j$, then the same local event will be represented by two different
projectors in the Hilbert spaces $\HS_j$ and $\HS'_j$.  We shall make the
plausible assumption that these two projectors lead to one and the same
Heisenberg projector when mapped via \eqref{eqn30} to the reference space
$\HS_r$ using the appropriate time development operators $T_{rj}$ and
$T'_{rj}$.

	Next we make the very important assumption that the dynamics embodied
in the collection of unitary time transformations $T_{jk}$ is local in the
sense that whenever $R_j$ and $R_k$ are two regions which are spacelike
separated (i.e., each point in $R_j$ is at a positive spacelike separation from
each point in $R_k$), and $P_j$ and $Q_k$ are projectors referring to physical
events (or properties) in $R_j$ and $R_k$, respectively, the corresponding
Heisenberg operators commute:
\begin{equation}
  \hat P_j \hat Q_k =\hat Q_k \hat P_j.
\label{eqn32}
\end{equation}
This is often referred to as the principle of causality \cite{Hg92}.  For our
analysis it has the important consequence that in cases in which more than one
time ordering is possible for a collection of regions $\{R_j\}$, because some
of the regions are spacelike with respect to each other --- for example, $R_0$
and $R_1$ in Fig.~\ref{fgr2} --- these different time orderings will give rise
to the same chain operators \eqref{eqn15}, since the corresponding Heisenberg
operators commute with each other.

	Suppose that $R_j$ consists of two or more disconnected subregions,
e.g., $R_3$ in Fig.~\ref{fgr2}(a).  We shall say that a projector $P_j$ which
is local to $R_j$ is in addition \emph{localized with respect to these
subregions} if it is a product of projectors, one (possibly the identity) for
each subregion, i.e., local to this subregion. Otherwise $P_j$ is
\emph{entangled} with respect to these subregions.  The distinction is
important, because, as we shall see later, one may wish to embed the subregions
in distinct nonintersecting hypersurfaces, as in Fig.~\ref{fgr2}(b), which are
part of a time-ordered collection.  If $P_j$ is localized, this construction
causes no difficulty, because each of the factors making up $P_j$ is itself a
local projector on the corresponding subregion, and the physical interpretation
of this projector does not depend on the hypersurface in which the subregion is
embedded.  But if $P_j$ is entangled, one cannot change the embedding by
placing the subregions (at least those among which $P_j$ is entangled) in
distinct hypersurfaces without violating the condition, fundamental to our
construction of relativistic histories, that each projector representing a
single event in a history be associated with a particular hypersurface in a
time-ordered collection of such surfaces.  This difference between localized
and entangled projectors will play a significant role in the later discussion
of quantum paradoxes.

	\subsection{Lorentz invariance} 
\label{sct3c}

	Lorentz invariance requires that the ``laws of physics'' be the same in
every Lorentz frame.  In the preceding analysis the whole discussion has been
carried out for a single Lorentz frame, let us call it $\LS$.  What should we
expect if we use a different Lorentz frame $\LS'$, thought of as a different
choice for a coordinate system?

	Each spacelike hypersurface $S_j$ should be thought of as consisting of
a definite collection of space-time points which is unchanged when the new
coordinate system $\LS'$ is adopted.  All that happens is that the quartet of
numbers $r=(t,x,y,z)$ representing a particular space-time point is replaced by
a new quartet $r'=(t',x',y',z')$.  The symbol $S'_j$ can be used to denote the
same collection of space-time points as $S_j$, but relabeled using
the new coordinates. If in $\LS$ the hypersurface $S_j$ is specified by an
equation 
\begin{equation}
  t=\tau_j(x,y,z),
\label{eqn33}
\end{equation}
then in $\LS'$ the same hypersurface, denoted by $S'_j$, will be specified in
the same manner, by setting $t'$ equal to a different function
$\tau'_j(x',y',z')$.

	Let us assume that there are well-defined rules based upon the function
$\tau$ for assigning a Hilbert $\HS_j$ to the surface $S_j$, and that these
rules do not depend upon the Lorentz frame.  Of course they will assign a
different Hilbert space $\HS'_j$ to $S'_j$ because $\tau'$ is not the same
function as $\tau$.  However, we can expect that $\HS'_j$ is related to $\HS_j$
by a unitary map (bijective isometry) $L_j$ which carries some $|\psi\rgl$ in
$\HS_j$ onto a $|\psi'\rgl$ in $\HS'_j$ representing the same physical
property.  Next assume that the unitary time transformation $T_{jk}$ mapping
$S_k$ to $S_j$ is determined in a unique way by the two functions $\tau_j$ and
$\tau_k$, by rules which do not depend upon the Lorentz frame once these
functions are given.  In the same way, $T'_{jk}$ mapping $S'_k$ to
$S'_j$ will
be determined by the functions $\tau'_j$ and $\tau'_k$.  The
Lorentz invariance of the dynamics is then expressed by the requirement
\begin{equation}
  T'_{jk} = L_j T_{jk} L_k\ad
\label{eqn34}
\end{equation}
for every pair $j$ and $k$. 

	A history embodying the same physical events as in (\ref{eqn2}) will
when expressed using the Hilbert spaces $\HS'_j$ be of the form
\begin{equation}
  {Y'}^\al = {P'_0}^{\al_0}\od {P'_1}^{\al_1}\od {P'_2}^{\al_2}\od\cdots 
  \od {P'_f}^{\al_f},
\label{eqn35}
\end{equation} 
with 
\begin{equation}
  {P'_j}^{\al_j} = L_j P^{\al_j}_j L_j\ad.
\label{eqn36}
\end{equation}
It is then easy to show that the weights calculated using the chain operators
(\ref{eqn15}) for such histories in $\LS'$ are the same as their counterparts
in $\LS$, and the consistency conditions (\ref{eqn17}) hold in $\LS'$ if and
only if they hold in $\LS$, using the operator inner product defined in
(\ref{eqn11}), or the one in \eqref{eqn31}, provided $\hat\rho$ is replaced by
a suitable $\hat\rho'$.  Thus the descriptions in the two Lorentz frames are
physically equivalent to each other.  A final point has to do with locality and
the condition \eqref{eqn32} for Heisenberg operators associated with regions
which are spacelike separated from each other.  All one needs to note is that
regions which are spacelike separated in one Lorentz frame are also spacelike
separated in any other, and the transformation rules in \eqref{eqn36}
ensure that $\hat P_j\hat Q_k$ is identical to $\hat Q_k\hat P_j$ if and only
if $\hat P'_j\hat Q'_k$ is the same as $\hat Q'_k\hat P'_j$.

	To be sure, all the difficulties of Lorentz invariance have been
``buried'' in the assumption that appropriate transformations $L_j$ exist, and
that the unitary time transformations satisfy (\ref{eqn34}), whatever inertial
frame $\LS'$ is employed. This, however, is as it should be: the present paper
is not devoted to the difficult task of constructing a Lorentz-invariant
relativistic theory.  Instead, its purpose is to show how various quantum
paradoxes are to be resolved, by the appropriate use of histories, within the
framework of such a theory, assuming it exists.

	\section{Wave Function Collapse}
\label{sct4}

	\subsection{Introduction}
\label{sct4a}

	Imagine a particle emitted in a nuclear decay, moving outwards as a
spherical wave packet.  When detected by a detector some distance away, its
wave function, according to textbook quantum theory, collapses instantaneously
to zero everywhere outside the detector, since that is where the particle is
now located.  This collapse helps explain why the particle cannot be detected
later by a second detector located further from the original decay.  But the
notion of such a collapse has troubled many physicists ever since the earliest
days of quantum theory \cite{nnne}\vphantom{\cite{Jmmr74}}.  It is troubling
because, among other things, what is instantaneous in one Lorentz frame is not
instantaneous in another, and therefore in some Lorentz frames the collapse
will travel faster than the speed of light, or even backwards in time, placing
the effect earlier than the cause. In addition, if after a suitable time the
detector has \emph{not} detected the particle, the probability increases that
the particle will be detected by another detector located further away, unless
this second detector is shadowed by the first, so even nondetection can alter
(collapse?)  the particle's wave function.  (This has led to the rather
confusing idea of an ``interaction-free'' measurement; see \cite{Dcke81} and
pp.~495f of \cite{Jmmr74}.)

\begin{figure}[h] 
$$
\begin{pspicture}(-6,-0.4)(6,1.0) 
\def\lwn{0.025} 
\def\lwd{0.04}  
\def\arxa{0.5} \def\arxb{0.6} \def\ary{1.0} 
\psset{
labelsep=2.0,
arrowsize=0.150 1,linewidth=\lwd}
\def\rectg(#1,#2,#3,#4){
\psframe[fillcolor=white,fillstyle=solid](#1,#2)(#3,#4)}
\def\wpack{
\pscurve(-1,0)(-.5,0.1)(0,.8)(.5,0.1)(1,0)}
\def\larrow{\psline{->}(\arxa,\ary)(-\arxb,\ary)}
\def\rarrow{\psline{->}(-\arxa,\ary)(\arxb,\ary)}
\rectg(-6,0,-5,1) \rput(-5.5,0.5){$A$}
\rectg(-1,0,0,1)  \rput(-0.5,0.5){$S$}
\rectg(5,0,6,1)   \rput(5.5,0.5){$B$}
\rput(-3,0){\wpack} \rput(-3,0){\larrow}
\rput(2,0){\wpack}  \rput(2,0){\rarrow}
\rput(-3,-0.4){$|\phi_a\rgl$}
\rput(2,-0.4){$|\phi_b\rgl$}
\end{pspicture}
$$
\caption{%
Wave packets representing a single particle, moving left and right from a
source $S$ towards detectors $A$ and $B$.}
\label{fgr3}
\end{figure}
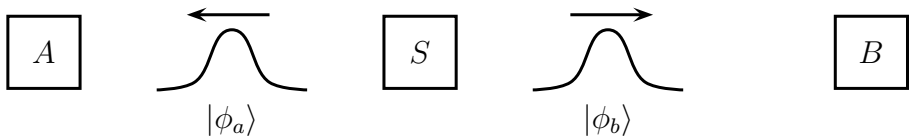

	In order to focus on essentials and simplify the discussion of how
a histories approach resolves (or tames) these problems, it is useful to
consider the analogous situation in one spatial dimension, as shown in
Fig.~\ref{fgr3}, where the wave function of the particle (\emph{one}
particle, not two!) is given by a linear superposition
\begin{equation}
  |\psi(t)\rgl = \blp |\phi_a(t)\rgl + |\phi_b(t)\rgl\brp/\st
\label{eqn37}
\end{equation}
of two wave packets moving outwards from a central source $S$ towards two
detectors $A$ and $B$, with $A$ closer to $S$ than $B$.  If $A$ detects the
particle, then at that instant of time (according to the collapse idea) the $b$
part of the wave packet in \eqref{eqn37} vanishes, whereas if $A$ does
\emph{not} detect the particle, the superposition \eqref{eqn37} is to be
instantly replaced by $|\phi_b(t)\rgl$.  Figure \ref{fgr3} is only schematic;
we are interested in situations in which the distances separating source
and detectors are very much larger than the widths of the wave packets, perhaps
large enough that it takes light a significant amount of time to travel from
$S$ to $A$ or $B$ \cite{nnnt}.

	\subsection{Without detectors}
\label{sct4b}

\begin{figure}[h] 
$$ 
\begin{pspicture}(-7.5,-1.0)(7.0,4.5) 
\def\lwn{0.025} 
\def\lwb{0.05}  
\def\lwd{0.04}  
\psset{
labelsep=2.0,
arrowsize=0.150 1,linewidth=\lwd}
\def\thick{\psline[linewidth=0.1](-0.15,0)(0.15,0)}
		\def\figa{
\psline[linewidth=\lwn]{->}(-2,0)(2,0)
\psline[linewidth=\lwn]{->}(0,0)(0,4.2)
\rput[t](1.9,-.2){$x$}
\rput[r](-.2,4.1){$t$}
\psline[linewidth=\lwb,linestyle=dashed](0,0)(-2.1,4.2)
\psline[linewidth=\lwb,linestyle=dashed](0,0)(2.1,4.2)
\rput[b](-1.45,3.5){$a$}\rput[b](2.05,3.5){$b$}
\psline(-2,1.5)(2,1.5)
\psline(-2,3.0)(2,3.0)
\rput[r](-2.1,1.5){$t_1$}
\rput[r](-2.1,3.0){$t_2$}
\rput(0.0,-0.5){\textbf{(a)}}
} 
		\def\figb{
\psline[linewidth=\lwn]{->}(-2,0)(2,0)
\rput[t](1.9,-.2){$x$}
\psline[linewidth=\lwn]{->}(-2,-1)(2,1)
\rput[t](1.9,0.8){$x'$}
\psline[linewidth=\lwb,linestyle=dashed](0,0)(-2.1,4.2)
\psline[linewidth=\lwb,linestyle=dashed](0,0)(2.1,4.2)
\rput[b](-1.45,3.5){$a$}\rput[b](2.05,3.5){$b$}
\psline(-2,1.5)(2,1.5)
\psline(-2,3.0)(2,3.0)
\rput(-0.75,1.5){\thick}\rput(0.75,1.5){\thick}
\rput(-1.5,3.0){\thick}\rput(1.5,3.0){\thick}
\rput[r](-2.1,1.5){$t_1$}
\rput[r](-2.1,3.0){$t_2$}
\psline(-2,0.5)(2,2.5)
\psline(-2,2.0)(2,4.0)
\rput[r](-2.1,0.5){$t'_1$}
\rput[r](-2.1,2.0){$t'_2$}
\rput(0.0,-0.5){\textbf{(b)}}
} 
		\def\figc{
\psline[linewidth=\lwn]{->}(-2,0)(2,0)
\rput[t](1.9,-.2){$x$}
\psline[linewidth=\lwn]{->}(-2,-1)(2,1)
\rput[t](1.9,0.8){$x'$}
\psline[linewidth=\lwb,linestyle=dashed](0,0)(-2.1,4.2)
\psline[linewidth=\lwb,linestyle=dashed](0,0)(2.1,4.2)
\rput[b](-1.45,3.5){$a$}\rput[b](2.05,3.5){$b$}
\psline(-2,1.5)(-0.6,1.5)
\psbezier(-0.6,1.5)(-.2,1.5)(1,2.2)(2,2.7)
\psbezier(-2,0.3)(-1,0.8)(0.2,1.5)(0.6,1.5)
\psline(0.6,1.5)(2,1.5)
\psline(-2,3.0)(-0.6,3.0)
\psbezier(-0.6,3.0)(-.2,3.0)(1,3.7)(2,4.2)
\psbezier(-2,1.8)(-1,2.3)(0.2,3.0)(0.6,3.0)
\psline(0.6,3.0)(2,3.0)
\rput(-0.75,1.5){\thick}\rput(0.75,1.5){\thick}
\rput(-1.5,3.0){\thick}\rput(1.5,3.0){\thick}
\psline(-2,0.5)(2,2.5)
\psline(-2,2.0)(2,4.0)
\rput(0.0,-0.5){\textbf{(c)}}
} 
\rput(-5,0){\figa} \rput(0,0){\figb} \rput(5,0){\figc}
\end{pspicture}
$$
\caption{ (a) Wave packet trajectories (dashed) and constant $t$ lines in
the $\LS$ space-time diagram. (b) Additional constant $t'$ lines for
Lorentz frame $\LS'$.  (c) Alternative hypersurfaces replacing the constant $t$
lines.}
\label{fgr4}
\end{figure}
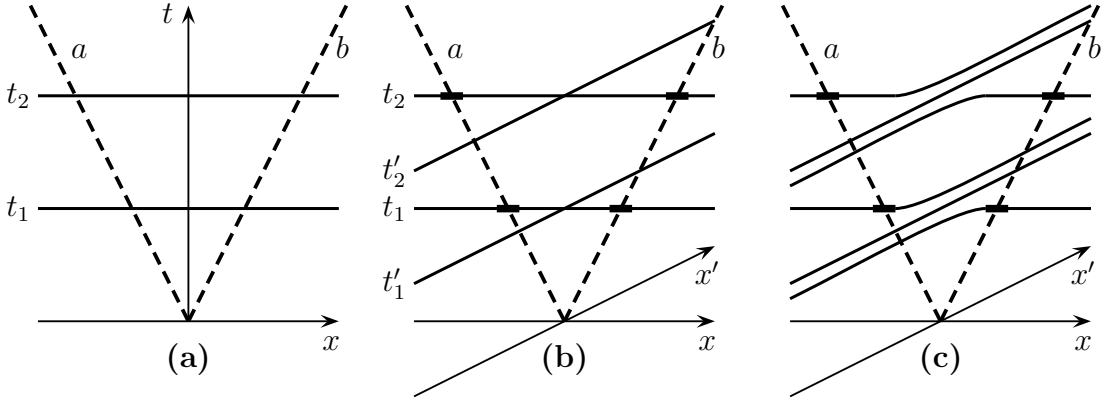

	As in Sec.~\ref{sct2d}, it is helpful to begin our analysis by
considering a situation in which there are no detectors present. The dashed
lines in Fig.~\ref{fgr4}(a) represent the centers of the wave packets
$|\phi_a(t)\rgl$ and $|\phi_b(t)\rgl$ in the Lorentz frame $\LS$ where their
velocities are equal and opposite.  Unitary time development then corresponds
to a family $\FS_0$ with support (as defined in Sec.~\ref{sct2d}) consisting of
the single history 
\begin{equation}
  \FS_0\coln\quad \psi(t_0)\od\psi(t_1)\od\psi(t_2)\od\cdots,
\label{eqn38}
\end{equation}
where $\psi(t)$ is the projector onto $|\psi(t)\rgl$.  To discuss
the location of the particle, and in particular whether it is to the left or to
the right of the source,  we introduce at
time $t_j$ a decomposition of the identity
\begin{equation}
  I_j = \sum_{\lm_j} P^{\lm_j}_j,
\label{eqn39}
\end{equation}
where the projectors $P^{\lm_j}_j$ project onto nonoverlapping intervals of the
$x$ axis chosen so that they are large in comparison to the widths of the
individual wave packets $\phi_a$ and $\phi_b$, but small compared to the
macroscopic length scales in Fig.~\ref{fgr4}.  They are also chosen so that
at each time $t_j$ both $\phi_a(t_j)$ and $\phi_b(t_j)$ are well inside one of
the intervals and not on the boundary between two of them.  This way the family
\begin{equation}
    \FS_1\coln\quad \psi(t_0)\od\{P^{\lm_1}_1\}\od\{P^{\lm_2}_2\}\od 
  \{P^{\lm_3}_3\}\od\cdots
\label{eqn40}
\end{equation}
will be consistent, and its support contains just two histories,
\begin{equation}
  \FS_1\coln\quad\psi(t_0)\od \left\{ \bem
 P^{a_1}_1\oda P^{a_2}_2\oda P^{a_3}_2\oda\cdots,
\\[\vmt]
 P^{b_1}_1\oda P^{b_2}_2\oda P^{b_3}_2\oda\cdots,
  \enm \right.
\label{eqn41}
\end{equation}
with equal weight.  The first history says that as time increases the
particle is in a series of intervals $a_1,\,a_2,\,\ldots$ falling along the
dashed line $a$ in Fig.~\ref{fgr4}(a), and the second that it is in a series of
intervals falling along $b$.  Thus they are coarse-grained quantum descriptions
that approximate classical trajectories. The two histories are mutually
exclusive possibilities: either the first occurs, so the particle follows
trajectory $a$, or the second, so the particle follows $b$.  The particle
cannot follow both trajectories, or hop from one to the other. If at time $t_2$
it is, say, in the interval $a_2$, then earlier it was in $a_1$, and later it
will be in $a_3$.

	The families $\FS_0$ and $\FS_1$ are incompatible, because $P^{a_j}_j$
and $P^{b_j}_j$ do not commute with $\psi(t_j)$, as follows from
\eqref{eqn37} and 
\begin{equation}
  P^{a_j}_j |\psi(t_j)\rgl = |\phi_a(t_j)\rgl/\st,\quad
  P^{b_j}_j |\psi(t_j)\rgl = |\phi_b(t_j)\rgl/\st.
\label{eqn42}
\end{equation}
To suppose that at $t_j$ the particle is in the physical state $\psi(t_j)$
\emph{and} that it is located in one of the two intervals $P^{a_j}_j$ or
$P^{b_j}_j$ is as meaningless as 
saying that a spin-half particle is in the state $S_z=+1/2$, and at the same
time ascribing to it values of $S_x$.

	Next consider a family $\FS_2$ with support consisting of
\begin{equation}
  \FS_2\coln\quad \psi(t_0)\od\psi(t_1)\od\left\{ \bem
 P^{a_2}_2\oda P^{a_3}_3\oda\cdots,
\\[\vmt]
 P^{b_2}_2\oda P^{b_3}_3\oda\cdots.
  \enm \right.
\label{eqn43}
\end{equation}
 Until $t_1$ the particle is in a nonlocal superposition, and thereafter it
either follows the (coarse-grained) $a$ trajectory or the $b$ trajectory, two
mutually exclusive possibilities, with probability 1/2.  One could, if one
wants to, say that the initial description in terms of $\psi(t)$ ``collapses''
between $t_1$ and $t_2$ onto another sort of description in which the particle
follows one of two distinct trajectories.  However, one should not think
of this ``collapse'' as a physical process. Instead it is the analog of a
description of a spin-half particle in terms of $S_z$ followed at a later time
in terms of $S_x$, as in \eqref{eqn22}.  The families
$\FS_0$, $\FS_1$, and $\FS_2$ are mutually incompatible in much the same way as
their counterparts in Sec.~\ref{sct2d}.  Each is a valid way of describing the
quantum particle, and there is no ``law of nature'' that specifies that one of
them is the ``correct'' description. However, there is a law of mathematics
which prevents one from combining them, since there is no way of representing
in the quantum Hilbert space a combination of events corresponding to
noncommuting projectors.

	There are always many incompatible ways of describing a quantum system,
and the choice among them depends on what one wants to discuss.  The use of
$\FS_2$ makes it possible to ascribe at time $t_1$ a relative phase to the sum
of the wave packets making up $|\psi\rgl$, in the sense that a $+$ sign occurs
rather than a $-$ sign on the right side of \eqref{eqn37}, but does not allow
one to assign a position to the particle, even to the extent of saying that it
is to the right or to the left of the source $S$.  Assigning a coarse-grained
position at this time requires that one use $\FS_1$, or something like it, in
which case the relative phase of the wave packets becomes a meaningless
concept.  Incidentally, once the ``split'' has occurred in the family $\FS_2$
the histories cannot be ``joined'' at a later time: replacing $P^{a_3}_3$ and
$P^{b_3}_3$ in \eqref{eqn43} with $\psi(t_3)$ violates the consistency
conditions, and the same comment applies to $\FS_1$.  In these respects the
situation is analogous to that of the spin-half particle considered in 
Sec.~\ref{sct2d}.

	\subsection{Different Lorentz frames} 
\label{sct4c}

	Consider a Lorentz frame $\LS'$ moving with respect to the frame
$\LS$ we have employed thus far, with constant-time surfaces shown in
Fig.~\ref{fgr4}(b) superimposed on the space-time diagram of (a). Let
\begin{equation}
  |\psi'(t')\rgl = \blp |\phi'_a(t')\rgl + |\phi'_b(t')\rgl\brp/\st
\label{eqn44}
\end{equation}
represent the wave function as it develops unitarily in time in the new Hilbert
space.  The obvious analogs $\FS'_0$, $\FS'_1$, and $\FS'_2$ of the families
considered previously can be obtained by adding primes to the appropriate
symbols in \eqref{eqn38}, \eqref{eqn41}, and \eqref{eqn43}, and the remarks
made above about the physical interpretations of the $\FS_j$ apply equally to
the $\FS'_j$.  

	The three families $\FS'_0$, $\FS'_1$ and $\FS'_2$ are not only
incompatible with one another, each is also incompatible with each of the three
families $\FS_0$, $\FS_1$, and $\FS_2$, because the constant-time hyperplanes
of $\LS'$ intersect those of $\LS$, and there is no way of placing them in a
time-ordered sequence. However, the incompatibility of $\FS_1$ and $\FS'_1$ is
only apparent, and can be removed by employing the ``trick'' shown in
Fig.~\ref{fgr4}(c).  Here the finite regions, shown with heavy lines, where the
particle can be located at $t_1$ and $t_2$ in $\FS_1$ have been embedded into
an alternative set of hypersurfaces which do not intersect, and are thus
compatible with, the hyperplanes used in $\FS'_1$.  This construction is
possible, as indicated at the end of Sec.~\ref{sct3b}, provided we are
interested in properties which are localized in the separate subregions, rather
than entangled among them. In the family $\FS_1$ we are concerned with local
properties: whether the particle is located in the $a$ subregion or in the $b$
subregion, rather than $\psi(t_1)$ and $\psi(t_2)$, which occur in the unitary
family $\FS_0$, and are entangled between the $a$ and the $b$ subregions.

	But does $P^{a_1}_1$, whose physical interpretation is that ``the
particle is in the (small) interval $a_1$'', really represent a \emph{local}
property? There is a subtlety here, for in the lowbrow approach outlined in
Sec.~\ref{sct3b} a local property is represented by a projector on the
Hilbert space of a (macro)local region, times the identity on another
Hilbert space for the rest of the universe.  In a Hilbert space of one-particle
wave packets, $P^{a_1}_1$ is not of this form, because it tells us both that
the particle is in $a_1$ \emph{and} that it is not in some distant region;
i.e., this projector provides more than local information.  The way to get
around this is to employ a many-particle Hilbert space, define $P^{a_1}_1$ to
be the projector that tells us there is exactly one particle in $a_1$, and use
the initial state $\psi(t_0)$ to specify that the universe contains only one
particle, as well as giving the wave packet for this one particle. In a history
whose initial state is $\psi(t_0)$, and given a dynamical law that the particle
cannot disappear or other particles appear, the event $P^{a_1}_1$ will allow us
to infer that the particle is in $a_1$ and therefore not elsewhere, even though
the projector $P^{a_1}_1$ by itself provides only local information.  The
reader for whom this argument is unnecessary should ignore it, while he who
finds it inadequate is invited to construct a better version.

	We conclude that in terms of their actual physical contents, $\FS_1$
and $\FS'_1$ are compatible, with a common refinement, call it $\FS^*_1$, that
uses the time ordering associated with the collection of hypersurfaces in
Fig.~\ref{fgr4}(c).  The support of $\FS^*_1$ again consists of two histories,
one with the particle following trajectory $a$ in a coarse-grained sense,
described sometimes by an $\LS$ and sometimes by an $\LS'$ projector, and the
other following trajectory $b$ in a similar fashion.  Aside from the subtleties
associated with coarse graining, in both space and time, the trajectories agree
with the picture provided by classical physics, even though they arise from a
fully quantum-mechanical description.

	On the other hand, the trick just discussed cannot be used in order to
combine $\FS_1$ with $\FS'_2$.  While one can introduce a common set of
hypersurfaces as in Fig.~\ref{fgr4}(c), the common refinement will not satisfy
the consistency conditions.  The trouble is that the particle can be localized
on the $b$ trajectory at $t_1$ in $\FS_1$, and this precedes (in the time
ordering of the hypersurfaces) the entangled state between $a$ and $b$ at
$t'_1$ in $\FS'_2$.  As noted above, trying to ``uncollapse'' a quantum
description in this manner violates the consistency requirements.  It is like
introducing an $S_z$ description into \eqref{eqn21} after an $S_x$ description
has appeared.  In addition, one cannot combine $\FS_2$ with $\FS'_2$, for in
this case the events at $t_1$ in the former and at $t'_1$ in the latter are
both entangled, so the collection of hypersurfaces in Fig.~\ref{fgr4}(c) is no
longer of any use. (It is possible, see the comments near the beginning of
Sec.~\ref{sct7a}, that some consistent generalization of the rules given in
Sec.~\ref{sct3} might allow one to construct a common refinement in this case,
but possible extensions of these rules fall outside the scope of the present
paper.)

	\subsection{Detectors}
\label{sct4d}

	Most of the tools required to resolve (or tame) the paradox of wave
function collapse are now in hand; all that remains is to introduce
measurements.  This we do using a fully quantum-mechanical description of the
two particle detectors shown in Fig.~\ref{fgr3}.  Let their states when ready
to detect a particle be denoted by $|A(t)\rgl$ and $|B(t)\rgl$, respectively,
and suppose that the detection event for the particle when represented by wave
packet $|\phi_a\rgl$ is given by a unitary time development
\begin{equation}
 |\phi_a\rgl |A\rgl \ra |A^*\rgl.
\label{eqn45}
\end{equation}
Here $|A^*\rgl$ is a state in which this detector has detected the particle,
as indicated by the position of a large pointer, or some other macroscopic
change that clearly distinguishes it from the untriggered or ready state
$|A\rgl$.  The time arguments have been omitted in \eqref{eqn45}; one should
think of the left side as at a time $t'$ before the particle interacts with the
detector, while the right side as at a time $t''$ after the interaction, when
the particle is trapped inside the detector.  If, on the other hand, the
particle is represented by wave packet $|\phi_b\rgl$, it will not interact with
detector $A$, and the counterpart of \eqref{eqn45} is
\begin{equation}
 |\phi_b\rgl |A\rgl \ra |\phi_b\rgl |A\rgl.
\label{eqn46}
\end{equation}
The analogous expressions for the $B$ detector are:
\begin{equation}
 |\phi_a\rgl |B\rgl \ra |\phi_a\rgl |B\rgl,\quad
 |\phi_b\rgl |B\rgl \ra |B^*\rgl. 
\label{eqn47}
\end{equation}

\begin{figure}[h]
$$ 
\begin{pspicture}(-7.5,-0.7)(2.5,4.5) 
\def\lwn{0.025} 
\def\lwb{0.05}  
\def\lwd{0.04}  
\def\rdot{0.10} 
\psset{
labelsep=2.0,
arrowsize=0.150 1,linewidth=\lwd}
\def\thick{\psline[linewidth=0.1](-0.15,0)(0.15,0)}
\def\dot{\pscircle*(0,0){\rdot}}
		\def\figa{
\psline[linewidth=\lwn]{->}(-1.5,0)(2.5,0)
\rput[t](2.4,-.2){$x$}
\psline[linewidth=\lwn]{->}(0,0)(0,4.2)
\rput[r](-.2,4.1){$t$}
\psline[linewidth=\lwb,linestyle=dashed](0,0)(-1.25,2.5)
\psline[linewidth=\lwb,linestyle=dashed](0,0)(1.75,3.5)
\psline[linewidth=\lwb,linestyle=dashed](-1.25,0)(-1.25,4.2)
\psline[linewidth=\lwb,linestyle=dashed](1.75,0)(1.75,4.2)
\rput(-1.25,2.5){\dot}\rput(1.75,3.5){\dot}
\rput[t](-1.25,-0.1){$A$} \rput[t](1.75,-0.1){$B$} 
\rput[b](-0.6,0.5){$a$}\rput[b](0.6,0.5){$b$}
\psline(-1.5,1.5)(2.5,1.5)
\psline(-1.5,2.9)(2.5,2.9)
\rput[r](-1.6,1.5){$t_1$}
\rput[r](-1.6,2.9){$t_2$}
\rput(0.0,-0.5){\textbf{(a)}}
} 
		\def\figb{
\psline[linewidth=\lwn]{->}(-1.5,0)(2.5,0)
\rput[t](2.4,-.2){$x$}
\psline[linewidth=\lwb,linestyle=dashed](0,0)(-1.25,2.5)
\psline[linewidth=\lwb,linestyle=dashed](0,0)(1.75,3.5)
\psline[linewidth=\lwb,linestyle=dashed](-1.25,0)(-1.25,4.2)
\psline[linewidth=\lwb,linestyle=dashed](1.75,0)(1.75,4.2)
\rput(-1.25,2.5){\dot}\rput(1.75,3.5){\dot}
\rput[t](-1.25,-0.1){$A$} \rput[t](1.75,-0.1){$B$} 
\rput[b](-0.6,0.5){$a$}\rput[b](0.6,0.5){$b$}
\psline(-1.5,1.5)(2.5,1.5)
\psline(-1.5,2.9)(2.5,2.9)
\rput[r](-1.6,1.5){$t_1$}
\rput[r](-1.6,2.9){$t_2$}
\psline(-1.5,0.75)(2.5,2.75)
\psline(-1.5,2.15)(2.5,4.15)
\rput[r](-1.6,0.75){$t'_1$}
\rput[r](-1.6,2.15){$t'_2$}
\rput(0.0,-0.5){\textbf{(b)}}
} 

\rput(-5,0){\figa} \rput(0,0){\figb}
\end{pspicture}
$$
\caption{%
(a) Wave packet trajectories (sloping dashed lines) and detector trajectories
(vertical dashed lines) in the $\LS$ space-time diagram. (b) Additional
constant $t'$ lines for Lorentz frame $\LS'$.}
\label{fgr5}
\end{figure}
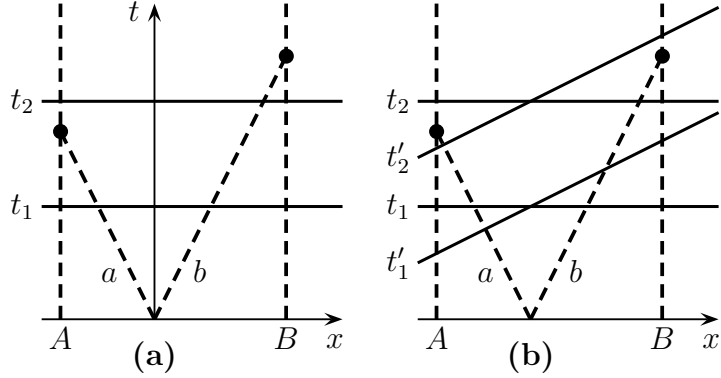
	
	With both detectors initially in the ready state, the overall unitary
time development corresponding to the space-time diagram in
Fig.~\ref{fgr5}(a) is represented by a family with support
\begin{equation}
  \GS_0\coln\quad \Psi_0\od \Psi_1\od \Psi_2\od \Psi_3\od \cdots,
\label{eqn48}
\end{equation}
where
\begin{align}
 |\Psi_0\rgl &= |\psi(t_0)\rgl |A\rgl |B\rgl 
\notag\\
 |\Psi_1\rgl &= |\psi(t_1)\rgl |A\rgl |B\rgl 
\notag\\
 |\Psi_2\rgl &= \blp |A^*\rgl + |\phi_b(t_2) |A\rgl\brp |B\rgl/\st
\notag\\
 |\Psi_3\rgl &= \blp |A^*\rgl|B\rgl + |A\rgl |B^*\rgl\brp/\st,
\label{eqn49}
\end{align}
and time arguments have again been omitted from the detector states.  Note that
since the detectors are macroscopic objects, both $|\Psi_2\rgl$ and
$|\Psi_3\rgl$ are examples of MQS states; see
\eqref{eqn26} and the comments following it.

	The family $\GS_1$ with support
\begin{equation}
  \GS_1\coln\quad \Psi_0\od\left\{ \bem
 P^{a_1}_1 AB \oda A^*B\oda A^*B \oda\cdots
\\[\vmt]
 P^{b_1}_1 AB \oda AB\oda AB^* \oda\cdots
  \enm \right.
\label{eqn50}
\end{equation}
is analogous to $\FS_1$ in Sec.~\ref{sct4b}.  In the first history the particle
follows the $a$ trajectory and is detected by $A$, while the $B$ detector is
unaffected. In the second history it is the $A$ detector that remains in its
ready state while the particle moves along trajectory $b$ and triggers $B$.
Note, in particular, that from the fact that detector $A$ triggers one can
conclude that the particle was earlier moving towards this detector, rather
than towards $B$, while if at $t_2$ detector $A$ has not detected the particle,
one can infer that the particle is (and was) moving towards detector $B$, and
will later be detected by $B$.  Such inferences are not at all mysterious, and
make no reference to wave function collapse.  Instead, they are consequences of
the fact that the two histories in \eqref{eqn50} are the only two
possibilities; all others have zero probability.  There are, to be sure, many
other frameworks that can be used to describe this situation in quantum terms,
but any framework that contains the events needed to draw the conclusions
stated above will assign them the same probabilities as $\GS_1$; see Sec.~16.3
of \cite{Grff02}.

	Another family $\GS_2$ with support 
\begin{equation}
  \GS_2\coln\quad \Psi_0\od\psi(t_1)AB\od\left\{ \bem
  A^*B\oda A^*B\oda\cdots 
\\[\vmt]
 \phi_b(t_2) AB\oda AB^* \oda\cdots
  \enm \right.
\label{eqn51}
\end{equation}
is analogous to $\FS_2$ in Sec.~\ref{sct4b} in that the particle remains in a
superposition state $\psi$ at time $t_1$, whereas at $t_2$ there has been a
``collapse'' into two possibilities: either the particle has been detected by
$A$, the first history, or, in the second history, it has not been detected by
$A$ and is still on its way to towards $B$, which will have detected it by
$t_3$.  (In the second history one could use the interval projector $P^{b_2}_2$
at time $t_2$ in place of the wave packet projector $\phi_b(t_2)$; for our
purposes it makes no difference.)  Note that just as the ``collapse'' in
$\FS_2$ is not a physical process, but represents a change in the type of
description being employed, so also in $\GS_2$ it is not something which is
brought about by some ``law of nature'', as is evident from the fact that
$\GS_0$ and $\GS_1$ are equally good descriptions, and in neither of them does
interaction with a measuring apparatus produce a corresponding ``collapse.''

	Introducing another description based on constant time (hyper)surfaces
in a second Lorentz frame $\LS'$, Fig.~\ref{fgr5}(b), leads to no new
principles beyond those already discussed in Sec.~\ref{sct4c}.  There is a
formal incompatibility between descriptions based upon constant $t$ and
constant $t'$ hyperplanes, but if one is concerned with local properties it is
possible to adopt a common refinement in which the particle either moves along
the $a$ trajectory to be detected by $A$, or along the $b$ trajectory to be
detected by $B$, and can be seen to do so using either $\LS$ or $\LS'$
projectors, provided that these descriptions are interleaved and one does not
try and impose them simultaneously at the same (macro)point in space-time.
Note, in particular, that if one is using such a local quantum description, the
fact that in $\LS'$ the particle can reach $B$ earlier
than it can reach $A$, the reverse from $\LS$, is no more paradoxical than
in classical relativistic physics.  It is only if one insists upon employing a
collapse picture using $\GS_2$, \eqref{eqn51}, along with its counterpart
$\GS'_2$ in $\LS'$, that difficulties arise.  These two families are
incompatible according to the rules of Sec.~\ref{sct3}, and it makes no sense
to ask which of them is correct, or when it is that the collapse ``really''
occurs, etc.

	\subsection{Summary}
\label{sct4e}

	It is useful to summarize the lessons provided by the preceding
analysis by restating its conclusions as they apply to the situation which
initiated our discussion: a particle moving outwards in a spherical wave, which
may later encounter a detector (or perhaps several detectors).  The spherical
wave corresponds to unitary time development (solving Schr\"odinger's
equation), and if unitary time development is applied to the full quantum
system of particle plus detector, the result will be an MQS state of a
triggered and untriggered detector.  While this, the analog of $\GS_0$ in
\eqref{eqn48}, is a perfectly valid quantum description, it is not useful for
answering questions such as: Did the detector detect the particle?  Where was
the particle before it was detected?  Posing these questions requires using
projectors which do not commute with the projector $\Psi(t)$ on the state
$|\Psi(t)\rgl$ resulting from unitary time evolution, and hence they are
meaningless within that framework. Instead, one must
use a family of stochastic histories in which at an appropriate time the
detector has or has not detected the particle, something analogous to $\GS_1$
or $\GS_2$ in \eqref{eqn50} and \eqref{eqn51}.  In families which are the
analogs of $\GS_1$, the particle follows a coarse-grained trajectory, the
quantum counterpart of a ``classical'' description, moving in a straight line
from the source of the decay until it reaches (in some histories) or misses (in
others) the detector. This is the type of description actually used by
physicists when thinking about decays of unstable particles, especially when
designing equipment with collimators and detectors, or considering sources of
undesirable background (see, e.g.,  pp.~123f in \cite{Omns94}). Because the
events in these families are \emph{local} in a coarse-grained sense, relative
to macroscopic length scales, their behavior under Lorentz transformations is
(essentially) the same as in classical relativistic physics.

	Wave function collapse is \emph{never} needed in order to produce
physically-meaningful quantum descriptions, since one can always assign
probabilities within a consistent family or framework using the Born rule and
its consistent extension, and then use these to calculate appropriate
conditional probabilities.  There are, to be sure, families of histories, the
analogs of $\GS_2$, which can be thought of as exhibiting a ``collapse''.
While these are perfectly legitimate quantum descriptions, the collapse can
occur in the absence as well as in the presence of a measurement, and
represents a change in the type of quantum description employed, not some sort
of physical process.  It is analogous to the physicist's choice to describe an
isolated spin-half particle during a certain time interval using $S_z$, and
during a subsequent time interval using $S_x$, even though the unitary
time development is trivial; see the comments following \eqref{eqn21}.

	\section{EPR Paradox}
\label{sct5}

	\subsection{Introduction}
\label{sct5a}

	The celebrated Einstein-Podolsky-Rosen \cite{EnPR35} or EPR paradox
is usually discussed nowadays using the formulation introduced by Bohm 
\cite{Bhm51ch22}
in which two spin-half particles $a$ and $b$ prepared in a spin-singlet state
\begin{equation}
  |s_0\rgl = \blp |z^+_a\rgl|z^-_b\rgl - |z^-_a\rgl|z^+_b\rgl \brp/\st,
\label{eqn52}
\end{equation}
where $|z^+_a\rgl$ is the state $S_{az}=+1/2$ of particle $a$, etc., fly apart
from each other, and the spin of one of the particles is later measured.  If
$S_{az}$ is measured and the outcome is $+1/2$, this means that $S_{bz}=-1/2$
for particle $b$, while an outcome of $-1/2$ implies that $S_{bz}=+1/2$.
Similarly, if $S_{ax}$ is measured, then $S_{bx}$ will have the opposite value:
$S_{bx}=-S_{ax}$.  The paradox is that one seems able to assign a value to
either $S_{bz}$ or to $S_{bx}$ depending upon which measurement is carried out
on particle $a$, and since the measurement should not influence particle $b$,
this seems to mean that both $S_{bz}$ and $S_{bx}$ have well-defined values,
contrary to the principles of quantum theory.

	Neither the original EPR formulation nor that of Bohm make use of
relativistic quantum theory.  But the paradox becomes a bit sharper in a
relativistic context, for particles $a$ and $b$ could be spacelike separated
when a measurement is made on $a$, so that any influence on $b$ would seem
contrary to the principles of relativity theory.  In addition, if the paradox
is formulated in terms of wave function collapse --- the spin state $|s_0\rgl$
changes instantly to either $|z^+_a\rgl|z^-_b\rgl$ or $|z^-_a\rgl|z^+_b\rgl$
when $S_{az}$ is measured --- one encounters the same problem noted in
Sec.~\ref{sct4a}: the collapse is not Lorentz invariant, as well as (or because
of) being instantaneous between spacelike separated points.

	Rather than a single measurement on particle $a$, one can imagine
separate spin measurements on $a$ and $b$, and if they are of the same
component, say $S_z$, then they will always give opposite results,
$S_{bz}=-S_{az}$. It is worth emphasizing that this sort of \emph{correlation},
even when the measurements are carried out in spacelike separated regions, is
not in itself paradoxical, as can be seen from a simple classical example.  A
pair of opaque envelopes is prepared, one containing a red and the other a
green slip of paper.  One envelope, chosen at random, is taken by astronaut
Alice on a voyage to Mars, while the other remains behind on the desk of Bob at
mission control.  By opening her envelope and observing (``measuring'') the
color of the slip of paper, Alice at once knows the color of the slip of paper
in Bob's envelope, and thus the color that Bob will observe (or perhaps has
already observed) when he opens it, even if that event occurs at a spacelike
separation.  As with every classical analogy, this one is not adequate for
illustrating all aspects of the quantum situation, but it does help clarify
what is and is not specific to quantum theory.

	\subsection{Without measurements}
\label{sct5b}

	As in Sec.~\ref{sct4}, we shall first analyze what happens in the
absence of measurements, assuming the two-particle wave function satisfying
Schr\"odinger's equation is given at time $t$ by
\begin{equation}
  |\psi(t)\rgl = |\om(t)\rgl |s_0\rgl,\quad 
  |\om(t)\rgl = |\phi_a(t)\rgl|\phi_b(t)\rgl,
\label{eqn53}
\end{equation}
where $|\phi_a(t)\rgl$ and $|\phi_b(t)\rgl$ are wave packets of the sort shown
in Fig.~\ref{fgr3}, except that now they refer to \emph{two distinct} (and
distinguishable) particles.  Their trajectories in a space-time diagram
are shown by the dashed lines 
 in Fig.~\ref{fgr4}, where once again
we assume that the distances are macroscopic, much larger than the
microscopic extent of a wave packet. 
Since we are interested in the spins rather than the positions of the
particles, it is convenient to ignore the latter, and think of 
\begin{equation}
  \FS_0\coln\quad \psi_0\od s_0\od s_0\od \cdots,
\label{eqn54}
\end{equation}
as a unitary history, with $\psi_0$ the projector on the initial state
$|\psi(t_0)\rgl$ and $s_0$ on the spin singlet state $|s_0\rgl$.  In this
family nothing can be said about any component of the spin angular momentum of
particle $a$ or of particle $b$, since the projectors for individual spin
states, such as $|z^+_a\rgl$, do not commute with $s_0$.

	More information about properties of individual spins is provided
by the family
\begin{equation}
  \FS_1\coln\quad \psi_0\od\left\{ \bem
 z^+_a z^-_b\od  z^+_a z^-_b\od \cdots,
\\[\vmt]
 z^-_a z^+_b\od z^-_a z^+_b\od \cdots,
  \enm \right.
\label{eqn55}
\end{equation}
where each history occurs with probability 1/2.  The physical interpretation is
straightforward: in the first history, particle $a$ has $S_{az}=+1/2$ and
particle $b$ has $S_{bz}=-1/2$ at all times later than $t_0$, whereas in the
second history $S_{az} = -1/2$ and $S_{bz} = +1/2$.  In either case the spins
are opposite, $S_{bz}=-S_{az}$, in the same way as the colors of the slips of
paper in the envelopes belonging to Alice and Bob.

	Still another consistent family
\begin{equation}
  \FS_2\coln\quad \psi_0\od s_0\od\left\{ \bem
 z^+_a z^-_b\od  z^+_a z^-_b\od \cdots\hfill
\\[\vmt]
 z^-_a z^+_b\od z^-_a z^+_b\od \cdots,
  \enm \right.
\label{eqn56}
\end{equation}
is analogous to $\FS_2$ in Sec.~\ref{sct4b}: up to $t_1$ the spins are in the
entangled singlet state, but thereafter they ``collapse'' into states in which
each particle has a well-defined value of $S_z$.  Of course this collapse, just
like those discussed in Secs.~\ref{sct2d} and \ref{sct4b}, has nothing to do
with any physical process, and instead reflects a change in the choice of basis
in which to describe the spins of the two particles; the comments following
\eqref{eqn43} apply equally in the present case.  The frameworks $\FS_0$,
$\FS_1$, and $\FS_2$ are mutually incompatible.  In addition, consistency
conditions mean that one cannot ``uncollapse'' the histories in $\FS_2$ (or in
$\FS_1$) by replacing the $S_z$ projectors at, say, $t_3$ with $s_0$; again,
the situation is analogous to that discussed in Sec.~\ref{sct4b}.

	There is nothing special about the $z$ direction.  The family
\begin{equation}
  \FS_3\coln\quad \psi_0\od\left\{ \bem
 x^+_a x^-_b\od  x^+_a x^-_b\od \cdots\hfill
\\[\vmt]
 x^-_a x^+_b\od x^-_a x^+_b\od \cdots,
  \enm \right.
\label{eqn57}
\end{equation}
is as good a quantum description as $\FS_1$, and replacing $z$ by $x$
everywhere in \eqref{eqn56} results in yet another consistent family.  All of
the frameworks discussed thus far are mutually incompatible, which does
\emph{not} mean that using one of them to construct a correct quantum
description of the particle's time evolution makes the others false, or that
one must invoke some hitherto unknown law of nature to decide which framework
is ``correct.''  Instead, think of each one as describing a somewhat different
``aspect'' of the time development of the quantum system, viewing it from a
somewhat different perspective, and thus each framework allows one to answer a
different set of physically sensible questions about the system.  How are the
values of $S_{ax}$ and $S_{bx}$ related to each other at some particular time?
This can only be answered by employing a framework in which the relevant
projectors occur at the time of interest; e.g., $\FS_3$ must be used rather
than $\FS_1$.

	One does not have to use the same component of spin angular
momentum for particles $a$ and $b$.  In the framework
\begin{equation}
   \FS_4\coln\quad \psi_0\od\left\{ \bem
 z^+_a x^+_b\od  z^+_a x^+_b\od \cdots,
\\[\vmt]
 z^+_a x^-_b\od  z^+_a x^-_b\od \cdots,
\\[\vmt]
 z^-_a x^+_b\od  z^-_a x^+_b\od \cdots,
\\[\vmt]
 z^-_a x^-_b\od  z^-_a x^-_b\od \cdots\hfill
  \enm \right.
\label{eqn58}
\end{equation}
the four histories occur with equal probability, and there is no correlation
between $S_{az}$ and $S_{bx}$.  For additional comments on this and other
examples, see Sec.~23.3 of \cite{Grff02}.


	\subsection{Measurements}
\label{sct5c}

	The spin measuring devices introduced in Sec.~\ref{sct2d} can also be
employed in the present context if supplied with a subscript to indicate which
particle is being measured.  For example, the device to measure $S_{az}$ has an
initial state $|Z_a\rgl$, and we assume that the unitary time development when
it interacts with particle $a$ has the form
\begin{equation}
  |z^+_a\rgl |Z_a\rgl \ra  |z^+_a\rgl |Z^+_a\rgl,\quad
  |z^-_a\rgl |Z_a\rgl \ra  |z^-_a\rgl |Z^-_a\rgl.
\label{eqn59}
\end{equation}
For an $S_{ax}$ measurement replace $Z$ with $X$ and $z$ with $x$.
Nondestructive measurements are not essential, but they simplify drawing
connections with traditional discussions using wave function collapse. 

	If we assume world lines as in Fig.~\ref{fgr5}(a), but with the $B$
detector eliminated, unitary time development starting with an initial state
\begin{equation}
  |\Psi_0\rgl = |\Psi(t_0)\rgl = |\om(t_0)\rgl|s_0\rgl |Z_a\rgl,  
\label{eqn60}
\end{equation}
in which the detector is ready to measure $S_{az}$, results in a succession of
states 
\begin{align}
 |\Psi(t_1)\rgl &= |\om(t_1)\rgl |s_0\rgl |Z_a\rgl,
\notag\\
 |\Psi(t_2)\rgl &= |\om(t_2)\rgl \blp |z^+_a\rgl |z^-_b\rgl |Z^+_a\rgl - 
	|z^-_a\rgl |z^+_b\rgl |Z^-_a\rgl \brp/\st,
\label{eqn61}
\end{align}
and so forth; for $t_3$ and all later times the spin and detector states are
the same as for $|\Psi(t_2)\rgl$.

	Now $|\Psi(t_2)\rgl$ is an
MQS state, so that the unitary family that contains it, the analog of $\GS_0$
in \eqref{eqn48}, cannot be used to discuss the outcomes of measurements.
Instead, we need something like
\begin{equation}
  \GS_1\coln\quad \Psi_0\od\left\{ \bem
 z^+_a z^-_b Z_a\od  z^+_a z^-_b Z^+_a \od  z^+_a z^-_b Z^+_a\od
  \cdots,
\\[\vmt]
 z^-_a z^+_b Z_a\od  z^-_a z^+_b Z^-_a \od  z^-_a z^+_b Z^-_a\od
  \cdots,
  \enm \right.
\label{eqn62}
\end{equation}
where the two histories occur with equal probability.  In the first of these
$S_{az}=+1/2$ and $S_{bz}=-1/2$ at times $t_1$ and later, and the measurement
outcome is $Z^+_a$ at times $t_2$ and later, as one would expect, while in the
other history $+$ and $-$ are interchanged.  This family corresponds to the
classical analogy introduced in Sec.~\ref{sct5a}, where astronaut Alice's
opening the envelope and seeing a red (or green) slip of paper reveals a prior
state of affairs, and enables her to conclude that the one in Bob's envelope is
of the opposite color.  Of course, this is not surprising given our earlier
discussion of the family $\GS_1$ in Sec.~\ref{sct2d} and $\GS_1$ in
Sec.~\ref{sct4d}.

	One can construct a family $\GS_2$, the analog of \eqref{eqn51},
in which the spin state in both histories is $|s_0\rgl$ at time $t_1$ and the
``collapse'' occurs in the same time step as the measurement:
\begin{equation}
  \GS_2\coln\quad \Psi_0\od s_0 Z_a\od \left\{ \bem
  z^+_a z^-_b Z^+_a \od  z^+_a z^-_b Z^+_a\od,
  \cdots,
\\[\vmt]
  z^-_a z^+_b Z^-_a \od  z^-_a z^+_b Z^-_a\od.
  \cdots,
  \enm \right.
\label{eqn63}
\end{equation}
An equally good family is
\begin{equation}
  \GS_4\coln\quad \Psi_0\od s_0 Z_a\od \left\{ \bem
  z^+_a x^+_b Z^+_a \oda  z^+_a x^+_b Z^+_a\oda\cdots,
\\[\vmt]
  z^+_a x^-_b Z^+_a \oda  z^+_a x^-_b Z^+_a\oda\cdots,
\\[\vmt]
  z^-_a x^+_b Z^-_a \oda  z^-_a x^+_b Z^-_a\oda\cdots,
\\[\vmt]
  z^-_a x^-_b Z^-_a \oda  z^-_a x^-_b Z^-_a\oda\cdots,
  \enm \right.
\label{eqn64}
\end{equation}  
the measurement counterpart of $\FS_4$, where the collapse occurs at the same
time, but now the properties of particle $b$ are uncorrelated with those of
particle $a$.  The existence of frameworks such as $\GS_1$ and $\GS_4$ as
alternatives to $\GS_2$ helps prevent one from drawing the erroneous conclusion
that a measurement carried out on particle $a$ has some mysterious long-range
influence on particle $b$.

	\subsection{Different Lorentz frames}
\label{sct5d}

	Consider a Lorenz frame $\LS'$ moving with respect to the frame
$\LS$ we have used up till now, with constant time ($t'$) surfaces as
shown in Fig.~\ref{fgr4}(b).  Frameworks $\FS'_j$ analogous to the $\FS_j$ of
Sec.~\ref{sct5b} can be defined by introducing primes on the appropriate
symbols in \eqref{eqn53} to \eqref{eqn58}, just as in Sec.~\ref{sct4c}, and
all comments made above on the physical interpretation 
of these families apply equally to these new descriptions.
	As in Sec.~\ref{sct4c}, each $\FS'_k$ is incompatible with each $\FS_j$
according to the rules of Sec.~\ref{sct3a}, but
in the case of $\FS'_1$ and $\FS_1$, which refer to \emph{local} properties,
one can use the ``trick'' in Fig.~\ref{fgr4}(c) in order to produce a common
refinement which includes the events of both frameworks for all $t_j$ and
$t'_j$ with $j>0$, with either $\psi_0$ or $\psi'_0$ (choose one or the other)
as the initial state.  That is, the relative time ordering of events with
spacelike separation is of no concern provided they are, indeed, spacelike
separated and not represented by entangled projectors, such as $s_0$. 

	There is, however, a complication not present in the earlier discussion
in Sec.~\ref{sct4c}, where we were only concerned with the presence or absence
of a particle in some region of space.  Here we are (at least potentially)
interested in different properties, always of the same particle, represented by
noncommuting projectors, such as $S_{az}$ and $S_{ax}$.  Suppose, for example,
we are interested in intercalating into the two histories in $\FS_1$ in
\eqref{eqn55} at some time between $t_1$ and $t_2$ a (local) property of
particle $a$.  If this is an $\LS$ event, in the sense of one defined using a
projector on a hyperplane which is at a constant time in $\LS$, then it must
satisfy the consistency conditions; in particular, if it is a projector onto a
spin state of particle $a$, it must be either $z^+_a$ or $z^-_a$.  If, instead,
we intercalate an $\LS'$ event, then it, too, must satisfy the consistency
conditions.  In either case these are determined, see \eqref{eqn17}, by
modified Heisenberg chain operators in which the additional event is
represented by its Heisenberg projector at an appropriate point in the defining
product \eqref{eqn15}.  That is to say, there are restrictions on which $\LS'$
properties can be consistently incorporated into an $\LS$ history, but they are
of precisely the same form governing the addition of $\LS$ events to that
history. While relativity theory adds technical complications, the basic rules
for consistency are exactly the same as in nonrelativistic quantum theory.

	When one is interested in \emph{nonlocal} properties represented by
projectors on entangled states between particles $a$ and $b$, then, as noted in
Sec.~\ref{sct4c}, the ``trick'' of introducing new hypersurfaces,
Fig.~\ref{fgr4}(c), will not work, and one must pay attention to the rules of
Sec.~\ref{sct3a} in order to avoid a situation in which one entangled state in
$\LS'$ ``occurs'' both before (for particle $a$) and after (for particle $b$)
another (entangled or product) state in $\LS$.  It is meaningless to combine
two such descriptions, in the precise sense that the theory as formulated in
Sec.~\ref{sct3} cannot assign a meaning to the combination, even though the
individual events are themselves parts of sensible quantum descriptions.

	Including measuring apparatus in the discussion leads to nothing new
beyond what has already been noted at the end of Sec.~\ref{sct4d}.  In
particular, if a measurement outcome is being used to infer a property of some
particle in a localized region, such an inference is possible whether or not
the particle is moving relative to the measuring apparatus. Of course, if one
is interested in a property of the particle in its own rest frame, this must be
appropriately related to the frame in which the calculation is carried out.
Such transformations, and their analogs in classical relativistic physics, are
not trivial, but these are technical issues not directly connected with the
paradoxes associated with wave function collapse.  The latter are best disposed
of by abandoning the notion of collapse, at least as some sort of physical
process, and instead using appropriate conditional probabilities based upon
histories.

	\section{Hardy's Paradox}
\label{sct6}

	\subsection{Statement of the paradox}
\label{sct6a}

	Hardy's paradox \cite{Hrdy92} resembles the EPR paradox in that it
involves two well-separated particles in an entangled state.  However, it is
more striking in that certain assumptions, including Lorentz invariance, seem
to lead to a contradiction: something is shown to be true that is known to be
false.  As well as the relativistic paradox discussed here, Hardy's original
paper contains a slightly different paradox whose discussion requires the use
of counterfactuals, and for that reason lies outside the scope of the present
paper.  Our exposition differs in some unimportant ways from Hardy's original,
and makes use of the nonrelativistic analysis in Ch.~25 of \cite{Grff02} (which
also discusses the counterfactual paradox).

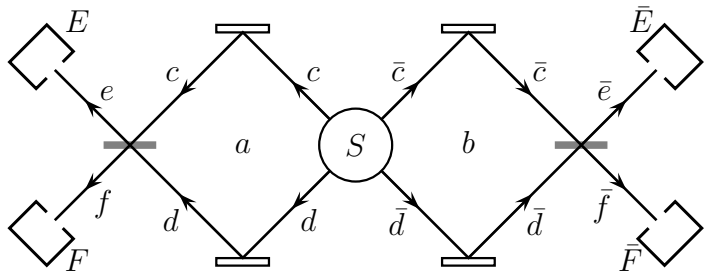
\begin{figure}[h]
$$
\begin{pspicture}(-4.7,-1.8)(4.7,1.9)
\psset{
arrowsize=0.150 1,linewidth=0.025,dash=0.1 0.1}
\def\lwp{0.035} 
\def\lwd{0.040} 
\def\lwm{0.030} 
\def\lws{0.030} 
\def\mrtp{0.05}\def\mrtn{-0.05}
\def\mrwp{0.35}\def\mrwn{-0.35}
\def\bstp{0.05}\def\bstn{-0.05}
\def\bswp{0.35}\def\bswn{-0.35}
\def\dtd{0.50}\def\dtdh{0.25} 
\def\dtwp{+0.35}\def\dtwn{-0.35} 
\def\dthp{+0.15}\def\dthn{-0.15} 
\def\srs{0.50}
	\def\mirror{
\pspolygon[linewidth=\lwm]%
(\mrwp,\mrtp)(\mrwn,\mrtp)(\mrwn,\mrtn)(\mrwp,\mrtn)
	}
	\def\bsplit{
\pspolygon[linestyle=none,fillstyle=solid,fillcolor=gray]%
(\bswp,\bstp)(\bswn,\bstp)(\bswn,\bstn)(\bswp,\bstn)
	}
	\def\detect{
\psline(0,\dthn)(0,\dtwn)(\dtd,\dtwn)%
(\dtd,\dtwp)(0,\dtwp)(0,\dthp)
	}
\def\deteb{\pscustom[linewidth=\lwd]{\rotate{45}\detect}}
\def\detfb{\pscustom[linewidth=\lwd]{\rotate{-45}\detect}}
\def\dete{\pscustom[linewidth=\lwd]{\rotate{135}\detect}}
\def\detf{\pscustom[linewidth=\lwd]{\rotate{-135}\detect}}
\rput(-3,0){\bsplit}\rput(+3,0){\bsplit}
\psline[linewidth=\lwp](0,0)(1.5,1.5)(3,0)(4,-1)
\psline[linewidth=\lwp](0,0)(1.5,-1.5)(3,0)(4,1)
\psline[linewidth=\lwp](0,0)(-1.5,1.5)(-3,0)(-4,-1)
\psline[linewidth=\lwp](0,0)(-1.5,-1.5)(-3,0)(-4,1)
\rput(-1.5,+1.5){\rput(0,\mrtp){\mirror}}
\rput(+1.5,+1.5){\rput(0,\mrtp){\mirror}}
\rput(-1.5,-1.5){\rput(0,\mrtn){\mirror}}
\rput(+1.5,-1.5){\rput(0,\mrtn){\mirror}}
\psline{->}(0,0)(0.85,0.85)
\psline{->}(1.5,1.5)(2.35,0.65)
\psline{->}(3,0)(3.6,0.6)
\psline{->}(0,0)(0.85,-0.85)
\psline{->}(1.5,-1.5)(2.35,-0.65)
\psline{->}(3,0)(3.6,-0.6)
\psline{->}(0,0)(-0.85,0.85)
\psline{->}(-1.5,1.5)(-2.35,0.65)
\psline{->}(-3,0)(-3.6,0.6)
\psline{->}(0,0)(-0.85,-0.85)
\psline{->}(-1.5,-1.5)(-2.35,-0.65)
\psline{->}(-3,0)(-3.6,-0.6)
\pscircle[linewidth=\lws,fillstyle=solid,fillcolor=white](0,0){\srs}
\rput(4,+1){\deteb}
\rput(4,-1){\detfb}
\rput(-4,+1){\dete}
\rput(-4,-1){\detf}
\rput[B](-1.5,-0.1){$a$}
\rput[B](+1.5,-0.1){$b$}
\rput(-2.25,0.75){\rput[Br](-0.10,0.10){$c$}}
\rput(0.75,0.75){\rput[Br](-0.10,0.10){$\bar c$}}
\rput(3.50,0.50){\rput[Br](-0.10,0.10){$\bar e$}}

\rput(-2.25,-0.75){\rput[tr](-0.10,-0.10){$d$}}
\rput(0.75,-0.75){\rput[tr](-0.10,-0.10){$\bar d$}}
\rput(3.50,-0.50){\rput[tr](-0.10,-0.10){$\bar f$}}

\rput(+2.25,0.75){\rput[Bl](0.10,0.10){$\bar c$}}
\rput(-0.75,0.75){\rput[Bl](0.10,0.10){$c$}}
\rput(-3.50,0.50){\rput[Bl](0.10,0.10){$e$}}

\rput(+2.25,-0.75){\rput[tl](+0.02,-0.10){$\bar d$}}
\rput(-0.75,-0.75){\rput[tl](+0.02,-0.10){$d$}}
\rput(-3.50,-0.50){\rput[tl](+0.02,-0.10){$f$}}
\rput(0,0){$S$}
\rput(-4,+1){\rput[B](0.3,0.5){$E$}}
\rput(+4,+1){\rput[B](-0.2,0.5){$\bar E$}}
\rput(-4,-1){\rput[B](0.30,-0.68){$F$}}
\rput(+4,-1){\rput[B](-0.35,-0.68){$\bar F$}}
\end{pspicture}
$$
\caption{%
Apparatus for Hardy's paradox; see text.}
\label{fgr6}
\end{figure}

	Imagine a source $S$, Fig.~\ref{fgr6}, that simultaneously
emits two particles $a$ and $b$ into the arms of two interferometers, in an
initial state
\begin{equation}
  |\psi_0\rgl = \blp |c\bar c\rgl +|c\bar d\rgl + |d\bar c\rgl\brp/\sqrt{3},
\label{eqn65}
\end{equation}
where $|c\bar c\rgl$  denotes a state in which
particle $a$ is moving through the $c$ arm of its interferometer on the left
side of the figure, and $b$ is moving through the $\bar c$ arm of the
interferometer on the right.  Note that \eqref{eqn65} has no $|d\bar d\rgl$
term, so it is never the case that $a$ is in the $d$ arm at the same time
that $b$ is in the $\bar d$ arm. 

\begin{figure}[h]
$$ 
\begin{pspicture}(-5,-0.7)(4.5,3.7) 
\def\lwn{0.025} 
\def\lwb{0.05}  
\def\lwd{0.04}  
\def\rdot{0.15} \def\rodot{0.20} 
\psset{
labelsep=2.0,
arrowsize=0.150 1,linewidth=\lwd}
\def\thick{\psline[linewidth=0.1](-0.15,0)(0.15,0)}
\def\dot{\pscircle*(0,0){\rdot}} 
\def\odot{\pscircle[fillcolor=white,fillstyle=solid](0,0){\rodot}} 
		\def\figa{
\psline[linewidth=\lwn]{->}(-2,0)(2,0)
\rput[t](1.9,-.2){$x$}
\psline[linewidth=\lwb,linestyle=dashed](0,0)(-1.5,3.0)
\psline[linewidth=\lwb,linestyle=dashed](0,0)(1.5,3.0)
\rput[r](-.5,0.7){$a$}\rput[l](.5,0.7){$b$}
\psline(-2,1.8)(2,1.8)
\psline(-2,3.0)(2,3.0)
\rput[r](-2.1,1.8){$t_1$}
\rput[r](-2.1,3.0){$t_2$}
\psline(-2,1.25)(2,3.25)
\psline(-2,3.25)(2,1.25)
\rput[r](-2.1,1.25){$t'_1$}
\rput[l](2.1,1.25){$t''_1$}
\rput(-0.9,1.8){\dot} \rput(0.9,1.8){\dot}
\rput(-1.5,3.0){\dot} \rput(1.5,3.0){\dot}
\rput(-1.2,2.4){\odot}\rput(1.2,2.4){\odot}
\rput[b](-1,1.10){$d_1$}\rput[b](1,1.10){$\bar d_1$}
\rput[b](-1.5,3.3){$E_2$}\rput[b](1.5,3.3){$\bar E_2$}
\rput(0.0,-0.5){\textbf{(a)}}
} 
		\def\figb{
\psline[linewidth=\lwn]{->}(-2,0)(2,0)
\rput[t](1.9,-.2){$x$}
\psline[linewidth=\lwb,linestyle=dashed](0,0)(-1.5,3.0)
\psline[linewidth=\lwb,linestyle=dashed](0,0)(1.5,3.0)
\rput[r](-.5,0.7){$a$}\rput[l](.5,0.7){$b$}
\psline(-2,1.25)(-0.9,1.8)
\psbezier(-0.9,1.8)(0,2.25)(1,1.5)(2,1.0)
\psbezier(-2,1.5)(-0.5,2.4)(0,2.25)(0.9,1.8)
\psline(0.9,1.8)(2,1.25)
\psline(-2,3.25)(-1.5,3.0)
\psbezier(-1.5,3.0)(0,2.25)(1.0,2.5)(2,3.0)
\psbezier(-2,3.5)(-1,3.0)(0.5,2.5)(1.5,3.0)
\psline(1.5,3.0)(2,3.25)
\rput(-0.9,1.8){\dot} \rput(0.9,1.8){\dot}
\rput(-1.5,3.0){\dot} \rput(1.5,3.0){\dot}
\rput(-1.2,2.4){\odot}\rput(1.2,2.4){\odot}
\rput(0.0,-0.5){\textbf{(b)}}
} 
\rput(-2.5,0){\figa} \rput(2.5,0){\figb}
\end{pspicture}
$$
\caption{Space-time diagram for Hardy's paradox. The open circles represent the
points where the particles pass through the beam splitters, the solid circles
at $t_2$ (in $\LS$) represent measurements which in $\LS'$ and $\LS''$ are
simultaneous with the corresponding points at $t_1$. In (b) these points are
on nonintersecting hypersurfaces.}
\label{fgr7}
\end{figure}
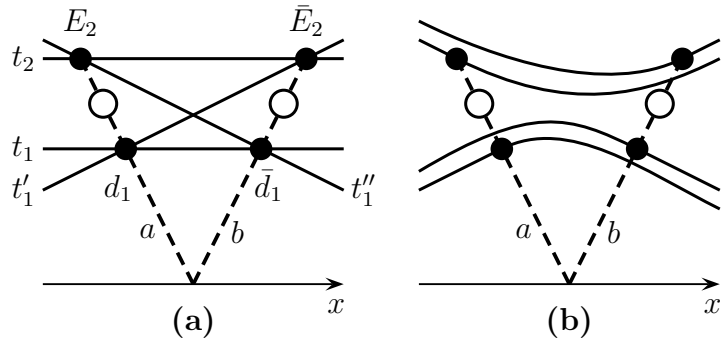

	The two beam splitters give rise to unitary time transformations
\begin{align}
  |c\rgl  \ra \blp |e\rgl + |f\rgl \brp/\st,\quad 
  &|d\rgl \ra \blp -|e\rgl + |f\rgl \brp/\st,
\label{eqn66}\\
  |\bar c\rgl \ra \blp |\bar e\rgl + |\bar f\rgl \brp/\st,\quad 
  &|\bar d\rgl \ra \blp -|\bar e\rgl + |\bar f\rgl \brp/\st,
\label{eqn67}
\end{align}
where for convenience we have chosen real phases (unlike \cite{Hrdy92}).
Unitary time development results in a state
\begin{equation}
  |\psi_2\rgl = 
 \blp -|e\bar e\rgl +|e\bar f\rgl + |f\bar e\rgl + 3|f\bar f\rgl\brp/\sqrt{12},
\label{eqn68}
\end{equation}
at a time $t_2$, Fig.~\ref{fgr7}(a), when both particles have passed through
the beam splitters.  Note that $|e\bar e\rgl$ occurs with a finite amplitude,
implying that $a$ and $b$ will be simultaneously detected by $E$ and
$\bar E$ with a probability of 1/12.

	If the interferometers are sufficiently large there will be a Lorentz
frame $\LS'$ in which particle $b$ is detected by $\bar E$ or $\bar F$ before
particle $a$ has reached the beam splitter on the left. It is then plausible
that just before detection occurs at the time $t'_1$ in $\LS'$, see
Fig.~\ref{fgr7}(a), the wave function for the two particles is obtained by
applying \eqref{eqn67} but not \eqref{eqn66} to \eqref{eqn65}, with the result
\begin{equation}
 |\psi'_1\rgl 
 = \blp 2 |c'\bar f'\rgl + |d'\bar f'\rgl + |d'\bar e'\rgl\brp/\sqrt{6},
\label{eqn69}
\end{equation}
where the primes indicate wave packets at constant time in $\LS'$.  From this
one can infer (e.g., by collapsing $|\psi'_1\rgl$ to $|d'\bar e'\rgl$) that if
$b$ is detected by $\bar E$, then at $t'_1$ particle $a$ is in the $d$ arm of
its interferometer.  Similarly, there will be a Lorentz frame $\LS''$ in which
$a$ is detected by $E$ or $F$ before $b$ reaches its beam splitter, and the
counterpart of \eqref{eqn69} is
\begin{equation}
 |\psi''_1\rgl 
 =\blp 2 |f''\bar c''\rgl + |f''\bar d''\rgl + |e''\bar d''\rgl\brp/\sqrt{6}.
\label{eqn70}
\end{equation}
From this it follows that if particle $a$ is detected by $E$, then at
$t''_1$ particle $b$ is in the $\bar d$ arm of its interferometer.

	Next assume that the presence of particle $a$ in arm $d$ at a point on
its trajectory indicated by $d_1$ in Fig.~\ref{fgr7}(a) does not depend upon
whether one describes it using $\LS$ or $\LS'$ or $\LS''$, and that there is a
similar invariance for particle $b$ relative to arm $\bar d$, and for which of
two detectors has detected a particle.  Hardy calls this assumption the
\emph{Lorentz invariance of elements of reality}, and it seems physically
plausible, especially if one thinks of extremely large interferometers, so that
the different Lorentz frames can be moving rather slowly with respect to each
other.  Assuming Lorentz invariance of this form, the inferences based on
\eqref{eqn69} and \eqref{eqn70} can be transferred to the Lorentz frame $\LS$,
and one arrives at the disquieting conclusion that in those cases (occurring
with probability 1/12) in which $a$ and $b$ are are simultaneously (in $\LS$)
detected in $E$ and $\bar E$ at time $t_2$, these particles were earlier, at
$t_1$, in the $d$ and $\bar d$ arms of their respective interferometers.  But
this conclusion is inconsistent with the initial state \eqref{eqn65}, since, as
noted previously, it lacks a $|d\bar d\rgl$ component.

	\subsection{Resolution of the paradox}

	It is helpful to analyze the logical structure of the argument leading
to the paradox in a bit more detail.  Using the space-time ``points'' (regions
small compared to the distance between beam splitters) labeled in
Fig.~\ref{fgr7}(a), the inferences based upon \eqref{eqn69} and \eqref{eqn70}
can be written in the form
\begin{equation}
 \bar E'_2 \Ra d'_1, \quad
 E''_2 \Ra \bar d''_1,
\label{eqn71}
\end{equation}
where $\bar E'_2$ means that in the Lorentz frame $\LS'$ particle $b$ has been
detected by $\bar E$ at $\LS$-time $t_2$, an event which in $\LS'$ is
simultaneous with the event $d'_1$: particle $a$ is in the $d$ arm of its
interferometer at $\LS$-time $t_1$. In a similar way, the $''$ events refer to
$\LS''$.  The assumption of Lorentz invariance of elements of reality implies
that these inferences are still valid if we add or delete primes from any of
the symbols in \eqref{eqn71}.  Combining the two inferences with primes
eliminated is what leads to the paradox.  What can one say about this in terms
of relativistic quantum histories?

	The inferences in \eqref{eqn71} refer to two hyperplanes which cross,
and therefore combining them is a violation of the single family rule as
formulated in Sec.~\ref{sct3}.  But, as already noted in Secs.~\ref{sct4c} and
\ref{sct5d}, one can get around this prohibition when considering local
properties, such as those in \eqref{eqn71}, by the device of introducing curved
hypersurfaces, as in Fig.~\ref{fgr7}(b).  What is essential is the time order
in which $d_1$ precedes $E_2$, and $\bar d_1$ precedes $\bar E_2$, which is
true in any Lorentz frame (e.g., $d''_1$ precedes $E''_2$), while the relative
temporal order of spacelike separated events, such as $d_1$ and $\bar E_2$, is
irrelevant, because we are not concerned with entangled states connecting two
spacelike separated regions.  To be sure, the wave functions in \eqref{eqn69}
and \eqref{eqn70} are entangled states in the sense just mentioned.  However,
their only role in the argument is that they are a way of calculating (via wave
function collapse) certain probabilities of local properties, probabilities
which could be calculated just as well by other methods which make no reference
to entangled states.  In the terminology of Sec.~9.4 of \cite{Grff02}, the
entangled wave functions in\eqref{eqn69} and \eqref{eqn70} are
pre-probabilities, and one need not think of them as representing physical
reality.  Thus each of the inferences in \eqref{eqn71} can be justified by
appeal to appropriate conditional probabilities, quite apart from the
mathematical method used to calculate the probabilities.

	Nevertheless, despite the fact that one is dealing with local
properties and hence the crossing of hyperplanes is of no concern, one can
only, as pointed out in Sec.~\ref{sct5d}, intercalate (local or nonlocal)
events at additional times into a quantum history if the consistency conditions
are satisfied.  This is a feature of both nonrelativistic and relativistic
quantum theory, and in the present instance it prevents one from combining the
two inferences in \eqref{eqn71}.  Each of these inferences is valid by itself,
in the sense that the events to the left and right of $\Ra$ can be placed in a
consistent family that confirms the correctness of the inference through
assigning a value of 1 to the corresponding conditional probability.  However,
the family required to justify the first inference is incompatible with that
required to justify the second, and the two cannot be combined, as one one
would have to do to reach a contradiction.

	To be more specific, any consistent history based on the initial state
$|\psi_0\rgl$ of \eqref{eqn65} which includes the event $d_1$ (or $d'_1$ or
$d''_1$) cannot also include the later event $E_2$ (or $E'_2$ or $E''_2$).
That is, it makes no sense to say that particle $a$ is earlier in the $d$ arm
of its interferometer and later detected by $E$. And what is meaningless --- an
``element of unreality'' --- in one Lorentz frame is equally meaningless in
another.  Each inference in \eqref{eqn71} refers to events which are spacelike
separated, so their Heisenberg projectors commute, and for this reason they are
compatible with the consistency conditions.  However, the conclusion of the
first inference is incompatible, in the quantum mechanical sense, with the
premise of the second inference, so putting the two together is not possible,
and this blocks the path to a logical contradiction.

	In summary, Hardy's relativistic paradox is resolved (or tamed) by
paying careful attention to using rules of reasoning that are compatible with
the mathematical structure of quantum theory.  In particular, chaining
arguments together in a manner which is perfectly acceptable in classical
physics cannot be done in the quantum context without first checking that they
belong to a single framework.  That is the basic lesson to be learned from the
Bell-Kochen-Specker result \cite{Mrmn93}, and from the extensive discussion of
quantum paradoxes in \cite{Grff02}.  Indeed, the procedure used here for
resolving the relativistic Hardy paradox is, in its essentials, identical to
that used for its nonrelativistic counterpart in Sec.~25.3 of \cite{Grff02}, to
which the reader is referred for additional details, including detailed
arguments for consistency and incompatibility of certain families.

	By contrast (and contrary to the conclusion of Hardy's original paper),
the assumption of Lorentz invariance of local elements of reality gives rise to
no problems: certain things one might expect to be the same in different
Lorentz frames, such as the presence or absence of a particle, are indeed the
same, or at least the assumption that this is true is not the origin of the
paradox.

	\section{Summary and Conclusions}
\label{sct7}

	\subsection{Relativistic histories}
\label{sct7a}

	The rules of nonrelativistic quantum kinematics summarized in
Sec.~\ref{sct2a} have a straightforward generalization to the relativistic
theory provided one adopts the condition, Sec.~\ref{sct3a}, that the spacelike
hypersurfaces used to build a relativistic family of histories be time ordered,
or, equivalently, cannot intersect each other.  In the nonrelativistic theory
the proper time ordering of events represented by Heisenberg projectors in the
product \eqref{eqn15} defining the chain operator is essential if one wants
physically reasonable results, and this seems to demand nonintersecting
hypersurfaces in the relativistic version, unless one wishes to construct an
entirely new theory.  However, if the Heisenberg operators associated with two
intersecting hypersurfaces commute with one another for the histories one is
interested in, the chain operator will not depend upon their order.  In
particular, this is true if the Heisenberg operators are identical, and that
suggests that combining certain unitary families (e.g., $\FS_0$ and $\FS'_0$ in
Sec.~\ref{sct4c}) may make sense.  Whether an extension of the rules of
Sec.~\ref{sct3a} allowing this sort of thing is worthwhile, and if so how best
to formulate it, are open questions.

	Locality and local properties are important concepts both for
formulating and for resolving quantum paradoxes. The approach in
Sec.~\ref{sct3b} seems adequate for the purposes of this paper, but could
undoubtedly be improved, especially by making its technical assumptions more
precise and less dependent on lowbrow intuition.  Part of this task is to give
a proper mathematical characterization of macrolocality, something which does
not look trivial given the difficulties associated with microlocality, as
mentioned in the introduction, though work by Omn\`es \cite{Omns97c} may be
pointing in the right direction.
	Another nontrivial task is that of constructing significant
Lorentz-invariant theories satisfying the conditions stated in
Sec.~\ref{sct3c}, in a way which can be applied to general hypersurfaces and
not just to hyperplanes.  The present paper contains nothing useful for this
task, unless it be a clarification of what it is that one is after.

	Such unresolved issues should not obscure the fact that the histories
approach extends in a very natural way from nonrelativistic to relativistic
quantum theory.  The basic formulas defining histories, Heisenberg chain
operators, weights or probabilities, and consistency conditions are formally
the same in the nonrelativistic approach as summarized in Sec.~\ref{sct2} and
in the relativistic extension in Sec.~\ref{sct3}. Not only are the symbols the
same, the associated concepts are extremely close if not completely
identical: the occurrence of events and histories, consistent families or
frameworks, refinements, incompatible frameworks, the single framework rule,
and probabilities.  Even the examples are similar.

	This close connection is hardly surprising given the fact, pointed out
in the introduction, that relativistic versions of the histories approach have
been around for some time. Nonetheless it is gratifying that some more recent
developments first formulated in a nonrelativistic context --- e.g.,
pre-probabilities and fully consistent schemes for assigning probabilities
based on different sorts of data --- can be  ``relativized'' without any
difficulty.  This straightforward compatibility stands in marked contrast to
the major difficulties which beset attempts to construct relativistic versions
in some other ``observer free'' quantum interpretations \cite{Drao99,Prle99}.

	\subsection{Resolving paradoxes}
\label{sct7b}

	The three paradoxes resolved, or at least tamed, in Secs.~\ref{sct4},
\ref{sct5} and \ref{sct6}, are all connected with the idea that a measurement
which takes place in some localized region can have effects at a distant place
spacelike separated from the region in question.  And they all invoke some form
of wave function collapse in order to calculate probabilities or make
inferences about the state of affairs at this distant place.

	The basic strategy by which the histories approach disarms these
paradoxes is by getting rid of wave function collapse.  How to do this is shown
in detail for the example considered in Sec.~\ref{sct4}; see the summary in
Sec.~\ref{sct4e}.  The conclusion is that wave function collapse is not needed
in quantum theory, and that if it is used it should \emph{never} be thought of
as a physical effect produced by a measurement.  Because of its misleading
connotations it might be best to get rid of wave function collapse
altogether. There is nothing that can be calculated or (correctly) inferred
using collapse which cannot be calculated or inferred equally well using
conditional probabilities based on fundamental quantum principles that make no
reference to measurements or to collapse.  The histories approach can supply
physical descriptions that resemble those of collapse (the $\GS_2$ families of
Secs.~\ref{sct4d} or \ref{sct5c}), and which help explain why the use of
collapse as a calculational procedure yields correct answers.  But it also
supplies alternative descriptions (the $\FS_1$ and $\GS_1$ families in these
same sections) which are much more useful for thinking about measurements from
a physical point of view, because they show how a measurement outcome is
related to some property of the microscopic system before the measurement took
place.  It is, in fact, this latter type of description that experimental
physicists use for designing their equipment and analyzing their data.  It is
to be regretted that textbooks which include the rather unrealistic model of
nondestructive measurements going back to von Neumann lack the basic concepts
needed to understand, from a quantum perspective, the devices actually used in
practice.

	The use of families of the $\FS_1$ or $\GS_1$ type, with macrolocal
properties both before as well as after a measurement (if any) takes place, has
the further advantage that it simplifies the discussion of how the description
of a quantum system must be altered in a relativistic theory if one uses a
moving coordinate system.  For families of this type, with an appropriate
coarse graining in space and time, the quantum description becomes
``classical'' (as one would anticipate from the work of Gell-Mann and
Hartle\cite{GMHr90b,GMHr93}), and Lorentz transformations
of particle trajectories behave the same way as in classical relativistic
physics.  States which are entangled over macroscopic distances, such as the
pre-measurement properties in the $\GS_2$ families, are not as easy to analyze,
but the histories approach provides the tools needed for using entangled
descriptions in a manner consistent with the basic principles of quantum
theory, or combining entangled and local states at different times in histories
in the same framework. 

	While there are many good reasons for removing wave function collapse
from nonrelativistic quantum mechanics, the case is even stronger for a
relativistic theory.  The use of collapse understood as some sort of physical
phenomenon is one of the main sources of the widespread notion that the quantum
world is inhabited by superluminal influences, leading to a prima facie
conflict with relativity theory.  It is then necessary to prove theorems to the
effect that these influences cannot carry information, i.e., they are
completely unobservable phenomena.  While a detailed discussion of the
(supposed) nonlocality of quantum theory lies outside the scope of the present
paper, it seems clear that to the extent that unobservable superluminal
influences arise from thinking of wave function collapse as a physical
phenomenon, disposing of the latter will get rid of the former.  In any case,
if collapse is not a physical phenomenon, discussions of when it
actually occurs \cite{HlKr70} are irrelevant to the physical theory.

	Once wave function collapse is out of the way (or has been tamed,
should one wish to continue using it), the resolution of the relativistic EPR
and Hardy paradoxes is fairly straightforward, using methods similar to those
used employed for their nonrelativistic counterparts in Chs.~23 to 25 of
\cite{Grff02}.  As long as one limits oneself to a single framework, there is
nothing paradoxical about EPR correlations in and of themselves, for they have
a simple classical analog, Sec.~\ref{sct5a}.  The notion that a measurement on
particle $a$ somehow influences particle $b$ can be effectively undermined by
noting some of the different frameworks that provide equally valid descriptions
of the quantum time development.
	In the case of Hardy's relativistic paradox the source of the
difficulty is not a failure of the Lorentz invariance of elements of reality,
such as the presence or absence of a particle in a given region of space, but
instead a process of reasoning which combines results from incompatible
frameworks.  In particular, the problem has to do with what one can meaningful
say about the time dependence of the state of a \emph{single} particle, rather
than measurements on a second particle spacelike separated from the first, and
thus relativistic considerations are actually irrelevant to the fundamental
conceptual difficulty.  Classical modes of reasoning easily give rise to
contradictions if imported into the quantum domain without regard to the way in
which the mathematics of quantum theory differs from that of classical
physics. 

	We believe that the paradoxes considered in this paper are
representative of a larger class, those in which traditional ideas of
measurement and wave function collapse give rise to contradictions, or to
nonlocal influences in apparent conflict with relativity theory.  If that is
true, then the methods used here for resolving the paradoxes of wave function
collapse, EPR, and Hardy should work equally well for this larger collection,
and help assuage the concern, seemingly widespread in the quantum foundations
community, that quantum theory and relativity are fundamentally incompatible.
The analysis in this paper indicates that the two go together very well when
proper account is taken of the rules which are needed to make even
nonrelativistic quantum mechanics a consistent theory.
	
	There remain, of course, the problems of microlocality, understanding
the quantum vacuum, constructing field theories using honest mathematics, and
the like, whose resolution is not brought any nearer by anything in this paper.
Unless it be indirectly through allowing a redirection of intellectual
energy away from enigmas whose ultimate origin is the unsatisfactory manner in
which probabilities have traditionally been introduced into quantum theory,
both nonrelativistic and relativistic, and which disappear when this is done in
a fully consistent way.

	\section*{Acknowledgments}

	  This research has been
supported by the National Science Foundation Grant PHY 99-00755.


\begin{thebibliography}{10}
\bibitem{nnnt}
In order to simplify the discussion we are assuming that the time required for
  decay is much shorter than the time required for the particle to reach one of
  the detectors. This need not be the case. For example, a nucleus undergoing
  alpha decay may have a lifetime of several seconds or longer, and in this
  case the collapse picture leads to the odd conclusion that detection of the
  alpha particle produces an instantaneous effect at the decaying nucleus. The
  histories approach allows one to view such processes in a manner which seems
  physically more realistic; see Sec.~12.4 of \cite{Grff02}.
\end{thebibliography}
\end{document}